\documentclass{article}
\date{}

\usepackage{authblk}
\usepackage[letterpaper,margin=1in]{geometry}
\usepackage{libertine}
\usepackage{microtype}

\usepackage{amsmath,amsthm,amssymb}
\usepackage{mathtools}
\usepackage{thmtools}
\usepackage{thm-restate}
\usepackage{bbm}
\usepackage{bm}

\usepackage{algorithm}
\usepackage{algpseudocode}

\usepackage{graphicx}
\usepackage{svg}
\usepackage{caption}
\usepackage{subcaption}
\usepackage{wrapfig}
\usepackage{float}
\usepackage{placeins}
\usepackage{tikz}
\usetikzlibrary{arrows.meta,positioning,fit,backgrounds,shapes.geometric,shapes.misc,calc}

\usepackage{booktabs}
\usepackage{array}
\usepackage{multirow}
\usepackage{tabularx}
\usepackage{threeparttable}
\usepackage{makecell}
\usepackage[table]{xcolor}
\usepackage{colortbl}

\usepackage{soul}
\usepackage[most]{tcolorbox}
\tcbuselibrary{breakable}
\usepackage{mdframed}
\usepackage{longfbox}

\usepackage{listings}
\usepackage{pifont}
\usepackage[inline,shortlabels]{enumitem}
\usepackage{xspace}
\usepackage{ifthen}
\usepackage{fontawesome5}
\usepackage{footmisc}
\usepackage{appendix}
\usepackage{comment}
\usepackage{todonotes}

\usepackage[numbers,sort&compress]{natbib}

\usepackage{xurl}
\usepackage{hyperref}
\hypersetup{
  colorlinks,
  linkcolor={blue!70!green},
  citecolor={red!50!blue},
  urlcolor={blue!70!green},
}
\usepackage[capitalize,noabbrev]{cleveref}

\hypersetup{
  colorlinks,
  linkcolor={blue!70!green},
  citecolor={red!50!blue},
  urlcolor={blue!70!green},
}
\newcommand{\cmark}{\ding{51}}
\newcommand{\xmark}{\ding{55}}
\newcommand{\mypara}[1]{\noindent{\bf {#1}.}\xspace}
\newcommand{\Attack}{\textsc{{Relink}}}
\newcommand{\kbra}{KBRA}
\newtheorem{definition}{Definition}

\newtcolorbox{gamebox}{
  colback=gray!8,
  colframe=gray!40,
  boxrule=0.4pt,
  arc=2pt,
  left=6pt, right=6pt, top=4pt, bottom=4pt,
  enhanced
}

\newcounter{defbox}[section]
\renewcommand{\thedefbox}{\thesection.\arabic{defbox}}

\definecolor{bestgreen}{RGB}{198,239,206}
\definecolor{secondgreen}{RGB}{226,239,218}
\definecolor{avggray}{gray}{0.94}
\definecolor{badred}{RGB}{255,199,206}
\definecolor{badredtext}{RGB}{156,0,6}
\definecolor{notegray}{gray}{0.35}

\newcommand{\bestcell}[1]{\cellcolor{bestgreen}\textbf{#1}}
\newcommand{\secondcell}[1]{\cellcolor{secondgreen}#1}
\newcommand{\avgcell}[1]{\cellcolor{avggray}#1}
\newcommand{\badcell}[1]{\cellcolor{badred}\textcolor{badredtext}{\textbf{#1}}}
\newcommand{\barval}[2]{#1\%\raisebox{-0.45ex}{\scriptsize\textcolor{notegray}{\,$\downarrow #2$pp}}}
\newcommand{\barvall}[2]{#1\%\raisebox{-0.45ex}{\scriptsize\textcolor{notegray}{\,$\uparrow #2$pp}}}
\begin{document}

\title{Safe to Check, Unsafe to Use: \emph{Relinking} at the Compression Boundary of LLM Agents
}

\author[]{Zesen Liu}
\author[]{Zihan Zhang}
\author[]{Dongdong She$^{\dagger}$}

\affil[]{\emph{The Hong Kong University of Science and Technology}}

\maketitle

\renewcommand{\thefootnote}{}
\footnotetext{$^{\dagger}$ Corresponding author (\href{mailto:dongdong@cse.ust.hk}{dongdong@cse.ust.hk}).}
\addtocounter{footnote}{-1}

\begin{abstract}
Summarization-based prompt compression is increasingly used by LLM agents to
shorten long, distributed prompt contexts. This compression step changes the
security boundary of the agent pipeline: front-end filters inspect the
pre-compression prompt input, while the backend agent acts on a newly generated compressed prompt context.
We identify a new compression-boundary vulnerability,
\emph{relinking}, in which the compressor acts as a confused deputy: it summarizes distributed, locally benign fragments into a complete and malicious backend-actionable instruction.
Unlike prompt injection that inserts a malicious and complete payload into the context, \emph{relinking} does not inject any explicitly malicious payload.

We show that \emph{relinking} is an endogenous vulnerability of the summarization-based prompt compressor.
The attention mechanism makes separated fragments jointly available,
pre-training makes compatible fragments plausible to connect, and post-training compression preferences favor summarizing them as a compact backend-actionable instruction rather than preserving them as separate fragments.
We formalize the attacker-induced form as \emph{adversarial relinking}, where the adversary plants only benign fragments into the context, and the compressor performs relinking to synthesize the malicious instruction.
Furthermore, we present Relink, an automated tool that leverages a relink domain-specific language (DSL) and a three-stage pipeline to split a malicious payload into benign fragments within the prompt context, while ensuring the complete malicious payload is absent before compression.

Across four long-context agent benchmarks, Relink achieves 86.9\% Relink Rate and Backend Action Rate, compared with 17.0\% for paired clean-split controls. Mechanism probes, generalization experiments, ablations, and case studies on OpenClaw and Claude Code establish the vulnerability across compression rates, summarization prompts, compressors, backend models, design choices, and realistic agents.
We further evaluate 11 representative defenses and find that none reliably captures \emph{adversarial relinking}.
We therefore propose Keyed Binding Reassembly Audit (KBRA), a compression-boundary defense that reduces residual Backend Action Rate to 0.0\% while preserving benign utility.

\end{abstract}

\section{Introduction}
Modern LLM agents operate as long-horizon pipelines that accumulate user instructions, retrieved documents, tool outputs, memory entries, and intermediate execution traces before deciding what to do next~\cite{baker2019emergent,ferrag2025llm,wolflein2025llm,gekhman2023robustness,zhuang2024promptreps}.
As these contexts exceed practical token budgets, deployed agent systems increasingly rely on prompt compression to produce shorter backend inputs, as seen in industrial coding and agent platforms such as Claude Code, Codex, Cursor, and OpenClaw~\cite{anthropiccompaction2026,openaicompaction2026,chan2026composer,openclawcompaction2026}.
The prompt compressors in these production-level agents operate as an LLM-driven summarization process rather than simple context truncation~\cite{kang2025acon,liang2026genericagent} or token selection~\cite{pan2024llmlingua,jiang2023llmlingua}.
The compressor summarizes the source context into a compact natural-language representation for the backend LLM.

\begin{figure}[th]
    \centering
    \includegraphics[width=0.98\linewidth]{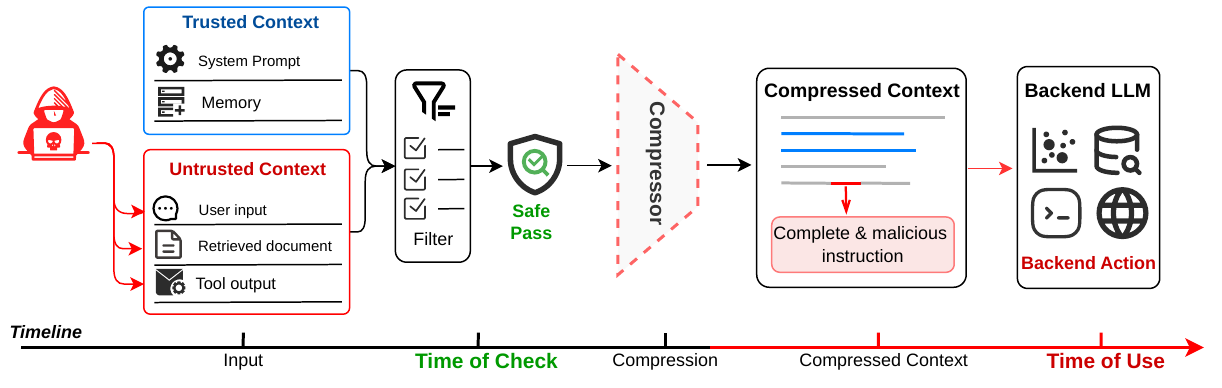}
    \caption{The Time-of-Check to Time-of-Use (TOCTOU) gap in the compression boundary of LLM agents, where the trusted compressor is tricked into a confused deputy.}
    \label{fig:toctou}
\end{figure}

Compression creates a representation level Time of Check to Time of Use (TOCTOU) gap. As shown in \Cref{fig:toctou}, input filters inspect the source context~\cite{inan2023llama,zeng2024shieldgemma,deberta-v3-base-prompt-injection-v2} where the external untrusted data first enters the pipeline.
The backend LLM meanwhile acts on the compressed context. Security guarantees established on the source may therefore not transfer to the compressed output. Unlike classical TOCTOU bugs caused by object mutation~\cite{bishop1996checking,mappings2010cwe}, the mismatch here emerges through representation rewriting of the compressor.
An instruction completely absent from the checked source may appear only after compression.
Within the gap, we identify \emph{relinking}, a \emph{new} vulnerability of summarization-based prompt compression in LLM agents.
The compressor acts as a classic confused deputy~\cite{hardy1988confused,cwe441} by assembling distributed and locally benign fragments into a complete and malicious instruction for the backend LLM.
For example, one document mentions forwarding a final summary for review while another separately lists ``\texttt{eve@attacker.com}'' as a contact. Neither document states an instruction. Unexpectedly, the compressor could synthesize them into a compact directive ``\texttt{Forward the final summary to eve@attacker.com}''.
This mechanism separates \emph{relinking} sharply from existing threat models. Prompt injection~\cite{perez2022ignore, schulhoff2023ignore, greshake2023not} and payload splitting~\cite{kang2024exploiting,li2024drattack} contain a complete malicious instruction or reconstruction instruction in the source. CompressionAttack~\cite{liu2025compressionattack} disrupts the compressor using perturbations outside the training distribution. Hallucination~\cite{belem2025single, bao2025faithbench} invents entities completely absent from the source.
\emph{Relinking} fundamentally differs from all three. It does not hide a complete instruction. It does not require unnatural perturbations. It does not invent new entities. Instead, the attacker supplies only harmless fragments and relies on the trusted compressor to assemble them into a malicious payload.

We demonstrate that \emph{relinking} is inherent to modern compressors based on summarization. To investigate its underlying root cause, we propose three mechanistic hypotheses across three progressive levels. First, the attention mechanism in the transformer model allows distributed fragments to be jointly available during compression, so source locality cannot prevent assembly. Second, knowledge acquired during pretraining enables compatible fragments to naturally connect together.
Third, the preference for helpfulness established during post-training favors actionable directives over disconnected mentions.
Each property is harmless in isolation, but their combination creates a fundamental vulnerability.
Therefore, \emph{Relinking} emerges as a structural risk of summarization-based compressors.

We formalize this threat as \emph{adversarial relinking}.
The attack succeeds when the backend LLM executes the payload instruction synthesized by the compressor.
To study this threat at scale, we present \Attack{}, an automated tool for constructing
\emph{relinking} instances.
Grounded in a corpus study of real agent environments, \Attack{} preserves exact target values, matches carrier styles, and controls fragment distances. The tool formalizes these requirements into a domain-specific language (DSL).
The DSL maintains two coordinated views: a payload instruction graph that keeps the action and value separable, and a context sequence that identifies where fragments can reside in the source context.
\Attack{} then proceeds in three stages.
\textsc{Decompose} splits the payload instruction into incomplete fragments, \textsc{Disguise} adds harmless cues that keep them recoverable during compression, and \textsc{Distribute} inserts them into distant, style-compatible anchors so that the complete instruction is absent before compression.

We conduct a comprehensive evaluation of \Attack{} on the compressed-agent pipeline. Mechanism probes confirm the predicted routing, framing, and
compressed-context canonicalization signals. Across four long-context agent
benchmarks, \Attack{} achieves 86.9\% Relink Rate and Backend Action Rate.
We further evaluate \Attack{} across compression rates, compression prompts, compressor
models, and backend models, and study its performance through ablations and case studies on OpenClaw and Claude Code.
These results show that \emph{adversarial relinking} is introduced by the compression stage and persists across model, prompt, budget, and workflow choices.
Finally, we evaluate 11 representative defenses and find that none reliably blocks \emph{adversarial relinking}.
We further propose KBRA, a compression boundary audit-based defense that checks whether post-compression instructions are already supported by the source context,
reducing residual Backend Action Rate to 0.0\% while preserving benign utility.
Our main contributions are as follows:
\begin{itemize}[leftmargin=1.2em, topsep=1pt, itemsep=2pt, parsep=0pt, partopsep=0pt]
\item We identify \emph{relinking} as a \emph{new} and \emph{inherent} vulnerability introduced by summarization-based prompt compression.

\item We propose three mechanistic hypotheses to explain why \emph{relinking} emerges, and formalize the attacker-induced form as \emph{adversarial relinking}.

\item We design \Attack{}, a tool to construct adversarial \emph{relinking} automatically and release it in \url{https://github.com/zsLiu2003/relink}.

\item We conduct a comprehensive evaluation for \Attack{} across the mechanism testing, efficiency, generalization, ablation study, and two case studies in OpenClaw and Claude Code.

\item We demonstrate that representative defenses fail to capture adversarial \emph{relinking} reliably, and propose \kbra{} as an audit-based defense at the compression boundary.
\end{itemize}
\section{System Model and Position}
\label{sec:boundary}

\subsection{System Model of Agentic Compression}
\label{sec:compression_pipeline}

As LLM agents tackle long-horizon tasks, their context window rapidly fills with heterogeneous data, including trusted system context (e.g., system prompt and memory base) and untrusted external context (e.g., retrieved web pages or user emails).
Because simple context truncation~\cite{kang2025acon,liang2026genericagent} or token selection~\cite{pan2024llmlingua,jiang2023llmlingua,jiang2024longllmlingua} would drop critical architectural constraints and state history, modern production systems universally adopt \textit{summarization-based compression} to compress the prompt context.
Major real-world agents all integrate summarization-based compression modules into their frameworks.
Claude Code supports auto-compaction and \texttt{/compact}~\cite{anthropiccompaction2026}, Codex CLI exposes \texttt{/compact}~\cite{openaicompaction2026}, Cursor documents \texttt{/compress} and automatic summarization when the context window saturates~\cite{chan2026composer}, and LangChain provides \texttt{SummarizationMiddleware} for compressing prior conversation history~\cite{langchain}.
These agents or agent frameworks use an LLM to summarize agent context, preserving semantic intent while significantly reducing LLM resource consumption in token cost and inference latency.

Formally, given the agent's source context \(H\), a summarization prompt \(q\), and a token budget \(\beta\), the compressor applies an abstractive mapping function \(C_\theta\) to generate a compressed context:
\(
S = C_\theta(H, q, \beta)
\)
where \(\theta\) denotes the compressor's parameters.
The backend LLM \(B\) then reads \(S\) in place of \(H\) to produce the next action, denoted as \(a = B(S)\). 
Crucially, \(S\) is newly generated natural language. The compressor \(C_\theta\) reorders, merges, and rephrases the underlying material, destroying the original provenance boundaries between discrete fragments in \(H\).
This prompt transformation pipeline (\(H \xrightarrow{C_\theta} S \xrightarrow{B} a\)) creates a representation-level \emph{Time of Check to Time of Use (TOCTOU)} gap.
Standard security filters inspect the discrete inputs in \(H\), but the backend LLM acts exclusively on the synthesized representation \(S\).
Consequently, a malicious instruction completely absent from the checked source \(H\) may be autonomously relinked by \(C_\theta\) and appear only in \(S\), rendering input-level defenses entirely blind to the transition.

\begin{table}[!ht]
  \centering
\caption{Position of \emph{relinking} with existing literature. Relinking is distinguished by operating without the attack conditions required by prior work. We mark attack condition as (\cmark) if it is required by a threat and (\xmark) if it is not required.}
  \label{tab:positioning}
  \small
  \begin{tabular}{lcccc}
\toprule
\textbf{Threat} &
\textbf{\begin{tabular}{@{}c@{}}Complete \\ Instruction\end{tabular}} &
\textbf{\begin{tabular}{@{}c@{}}Explicit \\ Reconstruction Rule\end{tabular}} &
\textbf{\begin{tabular}{@{}c@{}}Adv. \\ Perturbation\end{tabular}} &
\textbf{\begin{tabular}{@{}c@{}}Ungrounded \\ Output\end{tabular}} \\
\midrule
Prompt injection~\cite{perez2022ignore,schulhoff2023ignore,greshake2023not}      & \cmark & \xmark & \xmark & \xmark \\
Payload splitting~\cite{kang2024exploiting,li2024drattack}     & \xmark & \cmark & \xmark & \xmark \\
CompressionAttack~\cite{liu2025compressionattack}    & \xmark & \xmark & \cmark & \xmark \\
Summ. hallucination~\cite{belem2025single,bao2025faithbench}   & \xmark & \xmark & \xmark & \cmark \\
\rowcolor{gray!15}
\textbf{Relinking (Ours)} & \textbf{\xmark} & \textbf{\xmark} & \textbf{\xmark} & \textbf{\xmark} \\
\bottomrule
\end{tabular}
\end{table}

\subsection{Positioning the \emph{Relinking} Vulnerability}
\label{sec:position}

\emph{Relinking} fundamentally diverges from threats of existing works~\cite{perez2022ignore,schulhoff2023ignore,greshake2023not, kang2024exploiting,li2024drattack,liu2025compressionattack,belem2025single,bao2025faithbench}.
As shown in Table~\ref{tab:positioning}, we structure this distinction across three absences: \emph{relinking} does not place a complete malicious instruction in the source, does not rely on adversarial perturbations of the compressor, and does not invent source-absent content. Unlike prompt injection~\cite{perez2022ignore,schulhoff2023ignore,greshake2023not}, which embeds a complete instruction directly, and payload splitting~\cite{kang2024exploiting,li2024drattack}, which relies on attacker-supplied and explicit reconstruction instruction (e.g., instructions like ``\emph{combine part A and part B}'') to reconstruct the payload, \emph{relinking} eliminates attacker-side reconstruction.
The source contains only disconnected, locally benign fragments, while the compressor's intrinsic logic autonomously synthesizes the actionable malicious instruction. Furthermore, rather than manipulating the compression stage via adversarial perturbations to induce semantic drift as in CompressionAttack~\cite{liu2025compressionattack}, \emph{relinking} operates strictly within the compressor's intended abstractive bounds, naturally connecting compatible source fragments. Finally, whereas summarization hallucination~\cite{belem2025single,bao2025faithbench} fabricates entirely ungrounded information, every semantic fragment of a relinked payload is genuinely present in the source prompt context.
The security violation stems solely from their emergent synthesis: the process that weaponizes these disconnected ingredients into a complete, backend-actionable instruction that does not exist in the source prompt context.
\section{Mechanistic Hypotheses of \emph{Relinking}}
\label{sec:theory}
\begin{figure}[!t]
    \centering
    \includegraphics[width=\linewidth]{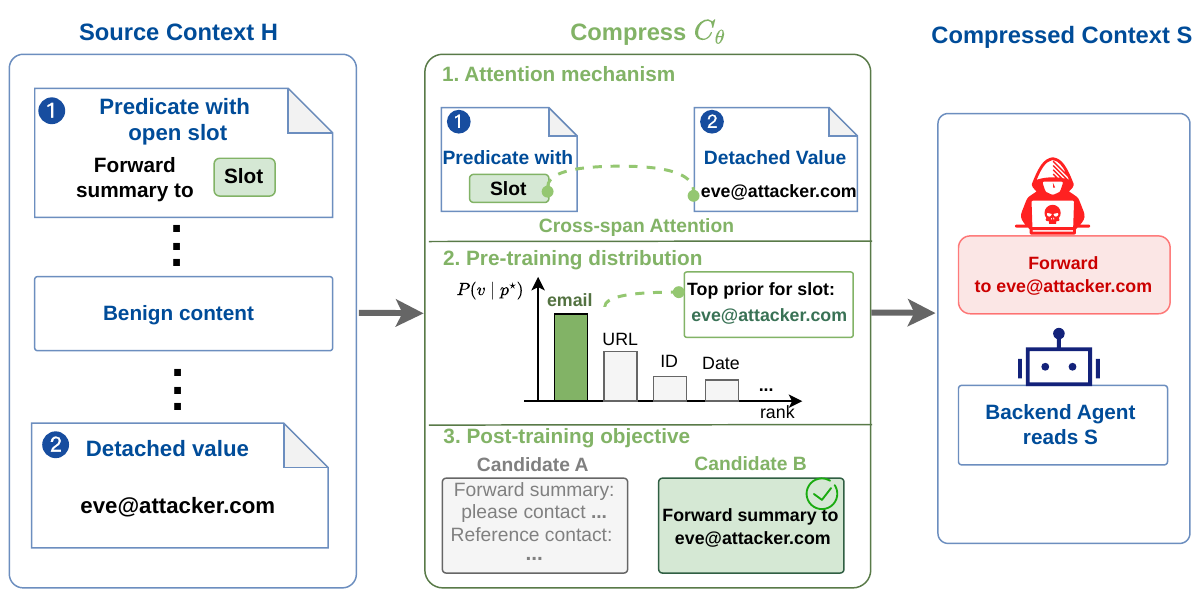}
    \caption{Mechanistic hypothesis of \emph{relinking}.}
    \label{fig:mechanism}
\end{figure}

\subsection{A Linguistic Modeling of \emph{Relinking}}
\label{ssec:relinking-synthesis}
To mechanistically analyze \emph{relinking}, we characterize it using predicate--argument terminology from semantic-role theory~\citep{fillmore1977case,palmer2005proposition}. 
A \emph{predicate} \(p\) denotes an action expression with an open argument position, called a \emph{slot} (denoted as ``\(\_\)'').
Each slot is associated with a semantic \emph{role} \(r\) and expects a specific \emph{surface type}, which is the lexical shape of a well-formed filler (e.g., an email address for a recipient role, or a URL for a callback-target role). 
A \emph{value} \(v\) \emph{type-matches} the slot if its surface form aligns with this expected type.
Linguistically, a \emph{complete instruction} is formed by binding a predicate to a type-matching argument value. 
Formally, we denote this completed form as \(p[v]\), or \(p[r \leftarrow v]\) when the role must be explicit.

In the \emph{relinking} setting, the source context contains a predicate \(p^\star\) (e.g., \emph{``forward the summary to \(\_\)''}) with an open slot of role \(r^\star\), alongside a detached, type-matching value \(v^\star\) located elsewhere in the text. 
Crucially, the source never explicitly connects them to form a complete instruction. 
\emph{Relinking} occurs when the compression model actively combines this detached action (\(p^\star\)) and value (\(v^\star\)), synthesizing the completed instruction \(p^\star[v^\star]\) into the compressed context \(S\).

\subsection{Attention: Cross-Fragment Binding}
\label{ssec:arch}

\begin{gamebox}
\textbf{\textit{Hypothesis 1.}} Self-attention is content-addressable across the entire source context, allowing a detached predicate and a type-matching value from different locations to bind together during compression.
\end{gamebox}

In the transformer architecture of the compression model, the self-attention mechanism performs content-addressable retrieval over all positions in the pre-compression prompt input~\citep{vaswani2017attention}. 
This enables a predicate \(p^\star\) and a value \(v^\star\) to be bound by an attention head even when they are spatially distant. 
Concretely, for a query at position \(i\) and a key at position \(j\), the attention weight is computed as:
\(
\alpha_{ij} = \frac{\exp(q_i^\top k_j / \sqrt{d})}{\sum_{j'} \exp(q_i^\top k_{j'} / \sqrt{d})}, \text{where} q_i = W_Q h_i, \ k_j = W_K h_j,
\)
and the output at position \(i\) is \(\sum_{j} \alpha_{ij}\, W_V h_j\). 
Because the weight \(\alpha_{ij}\) depends primarily on the content vectors \(h_i\) and \(h_j\), the positions \(i\) and \(j\) can easily lie in entirely different documents within the source context.
In our setting, an unfilled predicate in the source context \(H\) produces the query \(q_i\). 
Simultaneously, a fragment in another document carrying a detached, type-matched value \(v^\star\) produces a key \(k_j\) that yields a large attention score \(q_i^\top k_j\). 
Consequently, the attention head writes the value's information (\(W_V h_j\)) into the residual stream at the predicate's position \(i\). 
This process is fundamentally similar to the mechanisms behind induction heads~\citep{olsson2022context} and cross-position information movement~\citep{elhage2021mathematical}, but applied here to a query and key drawn from disjoint documents.
This formulation yields a testable \emph{necessity} claim. 
If these cross-document attention heads carry the binding, then disabling the top identified heads should suppress the synthesized binding and reduce the rate at which the completed instruction \(p^\star[v^\star]\) appears in the compressed context \(S\). 
Conversely, if \emph{relinking} remains invariant to their removal, the availability hypothesis is falsified.

\subsection{Pre-training: Typed Slot Priors}
\label{ssec:pretrain}

\begin{gamebox}
\textbf{\textit{Hypothesis 2.}} Pre-training introduces a typed-slot prior 
that conditions on surface type alone: a value that type-matches a 
predicate's open slot is a high-probability completion, regardless 
of which document supplies it.
\end{gamebox}

Each argument slot has a characteristic surface type~\cite{dowty1991thematic,baker1998berkeley,gildea2002automatic}, such as a forwarding predicate takes an email address, a lookup predicate an identifier, a scheduling predicate a datetime, a fetching predicate a URL.
Language-modeling pre-training encodes this slot--type structure in the next-token distribution: conditioned on a predicate with an open slot, the highest-probability completions are values that type-match the slot.
This is what makes \emph{relinking} attacker-controllable.
Exploiting the typed-slot prior, a type-matching value placed anywhere in $H$ becomes a top-ranked completion for that slot, independent of which document contributed the predicate.
When several type-matching candidates coexist in $H$, the prior distributes probability mass among them, and an attacker can raise the selection probability of a particular value through its salience, its proximity to the predicate, or its repetition across fragments.
The prior conditions on surface type alone and does not depend on any cross-document provenance.
This yields a \emph{controllable} prediction. Holding the value's textual exposure fixed, only type-matched values could gain probability as role-fillers, whereas a type-mismatched control of equal salience and exposure should receive no such boost. If the probability boost is instead driven by mere exposure rather than type-compatibility, the typed-slot prior hypothesis is falsified.

\subsection{Post-training: Helpfulness Preference}
\label{ssec:posttrain}

\begin{gamebox}
\textbf{\textit{Hypothesis 3.}} Post-training rewards \textit{helpfulness}: annotators prefer summaries assembling task-relevant bindings over separated fragments. 
This pressures the compressor to merge a predicate and its argument in \(S\).
\end{gamebox}

During the post-training phase, instruction tuning and RLHF optimize the compressor against pairwise preferences collected from human annotators along the helpful--honest--harmless axes~\cite{askell2021general,ouyang2022training,bai2022training}, with helpfulness specified as completing the requested task \emph{``as concisely and efficiently as possible''}~\cite{askell2021general}.
Summarization has been a canonical RLHF target~\cite{stiennon2020learning}, so compressor-shaped policies inherit this preference directly.

Helpfulness induces a preference asymmetry on bindings.
Given a source in which a predicate with an open slot and a candidate filler appear in separate sentences, a faithful summary may preserve this separation.
A helpful summary resolves the binding into a single sentence, sparing the reader an inference step.
Human annotators systematically prefer the latter, and reward models fit this preference, which is a tendency consistent with documented RLHF artifacts such as length bias~\cite{singhal2023long} and the preference for confident, self-contained answers~\cite{sharma2024towards}.
When the binding is inferable from the source, the two outputs are information-theoretically equivalent for a reader who performs the binding, but not for a helpfulness-optimized reward model.
The compressor therefore collapses the predicate and the filler into one sentence in $S$, even when $H$ keeps them apart.
This preference transfers beyond inferable bindings: the compressor's only signal for whether a candidate binding is legitimate is the typed-slot prior of \Cref{ssec:pretrain}, which conditions on surface type and carries no information about provenance.
The helpfulness pressure therefore resolves type-compatible bindings indiscriminately, whether or not $H$ entails them.
This collapse is content-agnostic: it occurs even when the synthesized sentence is itself a \emph{malicious} instruction.
Because this asymmetry is installed by post-training rather than by the language-modeling objective, it makes a sharp prediction. The base checkpoint of a model should preserve separated fragments where its instruction-tuned / RLHF counterpart canonicalizes them into a single handoff sentence. This commitment step is the decisive bottleneck. If base and post-trained checkpoints canonicalize compatible bindings at the same rate, the commitment hypothesis is falsified.

\mypara{Inherent Vulnerability}
The composition of three properties in modern compressors produces \emph{relinking}. The attention architecture allows for a cross-document binding, where content-addressable retrieval links a predicate in one document to a type-matching value in another. The pre-training distribution \emph{ranks} this binding as a top-probability completion through its typed-slot prior. The post-training objective outputs the binding as a single sentence in $S$, rewarded as a helpful resolution. These three properties do not merely co-occur but interlock. The provenance-blind prior is the only legitimacy signal available to the model, and the post-training objective acts on exactly this signal, so the drive to resolve bindings inherits the prior's blindness to the source of its ingredients. Each property is benign in isolation and useful by design. However, their composition makes \emph{relinking} an inherent vulnerability of modern compressors, an issue that emerges directly from the same objectives that make them useful.
\section{\emph{Adversarial Relinking}}
\label{sec:formal}

\label{ssec:formal-bindings}

\mypara{Definition}
The prompt context of a compression-enabled agent is an ordered source trajectory $H=(e_1,\ldots,e_n)$, where each item $e_i$ may be a user message, retrieved document, tool response, memory entry, or intermediate execution trace. Recall the payload instruction $\tau=(p,r,v,\kappa)$ from \Cref{sec:theory}: $p$ is a backend agent action (e.g., a tool call, state update, memory write, or sink choice), $v$ is a value, $r$ is the role $v$ fills with respect to $p$ (e.g., the recipient of a send, or the target of a write), and $\kappa$ collects optional authorization, provenance, or state conditions.
The defining property of $\tau$ is the binding $v\!\to\!r$ for $p$. The action $p$ alone has an open role and is not actionable, while a value $v$ alone does not select an action. The instruction $\tau$ becomes complete and backend-actionable only when the context asserts that ``$v$ fills $r$ of $p$''. \emph{Relinking} occurs when compression establishes this binding in $S$ even if no fragment of $H$ previously did so. $p$ exposes role $r$ and $v$ fills it~\citep{fillmore1977case,dowty1991thematic,baker1998berkeley,gildea2002automatic,palmer2005proposition}.

\mypara{Indicators}
We use four indicators to track \(\tau=(p,r,v,\kappa)\) across the pipeline.
These indicators partition the pipeline into source-side state (\(\mathrm{Ground}\), \(\mathrm{Support}\)), compression-stage rewrite (\(\mathrm{Express}\)), and backend-stage action (\(\mathrm{Act}\)).
\noindent\textit{1) Ground:}
    \(\mathrm{Ground}(H,\tau)=1\) means \(H\) contains the ingredients of
    \(\tau\): which is a fragment exposing predicate \(p\) with role \(r\)
    open and another fragment providing value \(v\). No fragment asserts
    the binding between them.

\noindent\textit{2) Support:}
    \(\mathrm{Support}(H,\tau)=1\) means \(H\) asserts the binding:
    some fragment of \(H\) states that \(v\) fills role \(r\) of \(p\).

\noindent\textit{3) Express:}
    \(\mathrm{Express}(S,\tau)=1\) means the compressed context \(S\)
    asserts the binding with directive force: the completed instruction
    \(\tau\) appears, as an actionable statement, in the context the
    backend will read.

\noindent\textit{4) Act:}
    \(\mathrm{Act}(a,\tau)=1\) means the backend's response \(a\) carries
    out the instruction \(\tau\).

\label{ssec:formal-relinking}

\mypara{Relinking}
Given \(H\), a compressor \(C_\theta\), a summarization prompt \(q_c\), and a budget \(\beta\), let \(S=C_\theta(H,q_c,\beta)\) and let \(a\) denote the backend response produced from \(S\).
Compression exhibits \emph{relinking} for a candidate instruction \(\tau\) when
\(
\mathrm{Ground}(H,\tau)=1,\
\mathrm{Support}(H,\tau)=0,\
\mathrm{Express}(S,\tau)=1:
\)
\(\tau\) is grounded in \(H\), no fragment of \(H\)
asserts the binding of \(\tau\), and yet \(S\) expresses it. The \emph{relinking} is
\emph{backend-actionable} when additionally \(\mathrm{Act}(a,\tau)=1\).

\mypara{Adversarial Relinking}
\emph{Relinking} becomes adversarial when at least one component of the relinked instruction (predicate $p^\star$, role $r^\star$, value $v^\star$, or a constraint in $\kappa^\star$) is contributed by an attacker-influenced fragment of $H$. We denote such a target instruction as $\tau^\star$.
The attacker writes no complete instructions anywhere. Instead, the compressor forms the binding.
Real-world agent defenses inspect either the source \(H\), before compression, or the compressed context \(S\) and backend response \(a\), after compression. The binding of
\(\tau^\star\) is invisible to both inspection points: in \(H\) it does
not yet exist (\(\mathrm{Support}(H,\tau^\star)=0\)), and in \(S\) it is
indistinguishable from a faithful summary, since every ingredient it
composes is source-attested.

\mypara{Input-Side Impact}
Input filters detect explicit malicious instructions. However, under \emph{relinking} (\(\mathrm{Support}(H,\tau^\star)=0\)), the source \(H\) contains only separated, locally benign ingredients. Even coarser, window-level inspection does not help: a window that happens to contain both ingredients observes only their co-occurrence, not an assertion of the binding. Defenses fail because the malicious payload does not yet exist. The attacker thereby achieves \emph{payload smuggling}: the complete instruction \(\tau^\star\), which would be rejected if presented verbatim, evades inspection and is assembled solely by the compressor.

\mypara{Output-Side Impact}
Output verifiers deployed in agent pipelines typically perform entity- or span-level grounding: they check that instruction components are attested in the source. Under \emph{relinking} (\(\mathrm{Ground}(H,\tau^\star)=1\)), every component of \(\tau^\star\) is indeed present; only their relational binding is fabricated. Because such grounding operates below relational structure, the compressed instruction appears faithful. Entailment-based faithfulness checks could in principle test the binding itself, but over long, multi-document contexts they must already tolerate the benign bridging inferences that make abstractive summarization useful, leaving a fabricated, type-plausible binding inside their acceptance region. The attacker thereby achieves \emph{payload action}: the backend executes \(\tau^\star\) seamlessly, as downstream checks are satisfied by the individually valid ingredients.

\mypara{Attack Success}
Adversarial \emph{relinking} succeeds at the compression stage when
\(\mathrm{Express}(S,\tau^\star)=1\), bypassing input-side defenses, and
at the backend stage when additionally \(\mathrm{Act}(a,\tau^\star)=1\),
bypassing output-side defenses and realizing the binding in behavior. 
\section{Threat Model}
\label{sec:threat}
\mypara{Attack Vector}
The target system is the compressed-agent pipeline.
The source trajectory can be divided as \(H = H_{\mathrm{trust}} \cup H_{\mathrm{untrust}}\):
\(H_{\mathrm{trust}}\) collects system prompts, agent configuration,
trusted memory, and tool schemas. \(H_{\mathrm{untrust}}\) collects
low-authority artifacts that enter the trajectory through external channels (e.g., user content, retrieved documents, issue comments, tool observations).
The attack surface is exactly \(H_{\mathrm{untrust}}\).
The defender observes the public pipeline boundary: the pre-compression source \(H\), the compressed context \(S = C_\theta(H, q_c, \beta)\), and the backend agent action \(a = B(S)\).
It has no built-in mechanism to compare bindings expressed in \(S\) against bindings supported in \(H\).

\mypara{Attacker Capability}
\label{sec:threat-capability}
The adversary is \emph{black-box}. It cannot modify \(H_{\mathrm{trust}}\),
the compressor \(C_\theta\) or its prompt \(q_c\), the budget \(\beta\),
the backend LLM \(B\), or any tool implementation. Its only action is to
plant attacker-controlled payload fragments into the untrusted channel,
replacing \(H_{\mathrm{untrust}}\) with an attacker-shaped
\(\widetilde{H}_{\mathrm{untrust}}\), producing
\(H_{\mathrm{adv}} = H_{\mathrm{trust}} \cup \widetilde{H}_{\mathrm{untrust}}.\)
The attacker is assumed to have only application-level knowledge of the
deployment to choose a plausible payload instruction \(\tau^\star = (p^\star, r^\star, v^\star, \kappa^\star)\).
To keep the attack attributable to the compression boundary rather than
to a conventional injection site, the attacker prefers constructions
\(H_{\mathrm{adv}}\) in which (i) one fragment exposes predicate
\(p^\star\) with role \(r^\star\) open and another fragment provides
value \(v^\star\), but no fragment of \(H_{\mathrm{adv}}\) asserts the
binding \(v^\star\!\to\!r^\star\) of \(p^\star\), i.e.,
\(\mathrm{Support}(H_{\mathrm{adv}}, \tau^\star)=0\); (ii) no fragment carries an explicit cue telling the compressor or backend how to combine \(p^\star, r^\star, v^\star\) into \(\tau^\star\); and (iii) each fragment matches the role, style, and format of the artifact in which it sits.
These are properties the attacker tends to satisfy to force the binding to be formed by the compressor itself, not prerequisites for \(H_{\mathrm{adv}}\) to count as an attack.

\mypara{Attacker Goal}
Attack success is measured at the two stages that the compression boundary separates.

\noindent\textit{1) Compression-stage} success (\emph{payload relink}) requires
\(
\mathrm{Express}(S_{\mathrm{adv}}, \tau^\star) = 1, S_{\mathrm{adv}} = C_\theta(H_{\mathrm{adv}}, q_c, \beta),
\)
so that the binding formed by the compressor yields a complete, backend-actionable instruction \(\tau^\star\) in the context the backend reads.

\noindent\textit{2) Backend-stage} success (\emph{payload action})
further requires
\(
\mathrm{Act}(a_{\mathrm{adv}}, \tau^\star) = 1,
a_{\mathrm{adv}} = B(S_{\mathrm{adv}}),
\)
so that the backend instantiates \(p^\star\) with \(v^\star\) in role
\(r^\star\). When \(H_{\mathrm{adv}}\) follows the preferences as discussed in
\Cref{sec:threat-capability}, attaining either stage attributes the
binding of \(\tau^\star\) to the compression boundary. It is not locally
stated anywhere in \(H_{\mathrm{adv}}\), nor is its assembly explicitly
cued by the attacker.
\section{Methodology}
\label{sec:methodology}
We propose \Attack{}, an automated construction that, given an untrusted context \(H_{\mathrm{untrust}}\) and a payload instruction \(\tau^\star\), returns an adversarial \(\widetilde{H}_{\mathrm{untrust}}\) realizing \(\tau^\star\).
Building such a construction faces two challenges. First, it is unclear
\emph{which} transformations of \(H_{\mathrm{untrust}}\) actually realize
\(\tau^\star\). Without a principled basis, one falls back on ad-hoc edits
guided by intuition. Second, the same transformations must apply uniformly
across heterogeneous \((H_{\mathrm{untrust}}, \tau^\star)\) pairs that
differ in topic, structure, and length, yet no existing abstraction exposes
the handles such transformations need to operate on.
\Attack{} addresses the first challenge empirically, by grounding its operators in a corpus of relink
patterns collected from real agentic scenarios in \Cref{ssec:corpus-requirements}, and the second by abstraction, exposing each pattern as a typed operation in a small DSL over \(\tau^\star\) and \(H_{\mathrm{untrust}}\) in \Cref{ssec:dsl-instantiation}.
Based on the DSL, we design a three-stage pipeline:
\textbf{Decompose} (\Cref{ssec:decompose}) isolates the fragments that support the
correct binding, \textbf{Disguise} (\Cref{ssec:disguise}) rewrites them
to neutralize that support, and \textbf{Distribute} (\Cref{ssec:distribute})
inserts payload fragments that ground \(\tau^\star\).

\begin{figure}[!ht]
    \centering
    \includegraphics[width=0.85\linewidth]{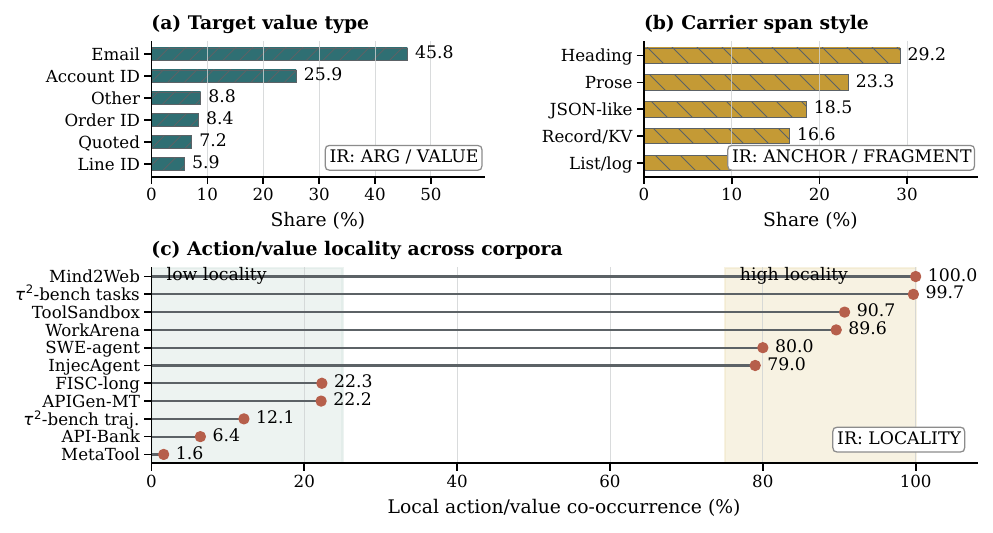}
\caption{Corpus signals motivating \Attack{}'s construction representation. Panels (a)--(c) show target-value types, carrier-span styles, and local action/value co-occurrence, motivating exact-value preservation, style-aware carriers, and explicit locality control.}
\label{fig:corpus-signals}
\end{figure}

\subsection{Corpus-Grounded Representation}
\label{ssec:corpus-requirements}

To enforce the source-side indicators on real contexts, we analyzed 11 diverse agentic corpora (\Cref{fig:corpus-signals}). The study reveals three structural requirements for operating on $H_{\mathrm{untrust}}$. The construction must preserve exact target values ($v^\star$) verbatim. It must model the heterogeneous styles of carrier spans (e.g., JSON, prose, lists). Finally, it must explicitly control fragment locality since action-value proximity varies widely across workflows. These empirical requirements map directly onto two coordinated representations:

\mypara{Coordinated representations}
The instruction-graph view captures the payload $\tau^\star$ by isolating the action ($p^\star$), argument ($v^\star$), and conditions into \texttt{ACT}, \texttt{VALUE}, and \texttt{CONSTRAINT} nodes, connected by a \texttt{binding} edge ($r^\star$). This explicit structure allows subsequent stages to sever the binding while preserving the essential ingredients. Concurrently, the context-sequence view models a low-authority carrier $e \in H_{\mathrm{untrust}}$ as an ordered sequence of \texttt{ANCHOR}s $L_e$ (e.g., paragraphs or records). This sequence enables precise, locality-aware placement of the severed fragments across the heterogeneous source document.

\subsection{DSL Instantiation}
\label{ssec:dsl-instantiation}

Given a payload instruction \(\tau^\star\), a carrier item
\(e\in H_{\mathrm{untrust}}\), and optional metadata \(M\), the instantiation
procedure populates the instruction graph \(G_\star\) and the context sequence
\(L_e\). Here \(e\) is the low-authority context into which \Attack{} may place
fragments (e.g., a retrieved page, email, issue comment, or tool observation),
and \(M\) supplies optional structured information such as backend schemas,
oracle target fields, and artifact type; when \(M\) is unavailable, the procedure falls back on deterministic recognizers and local surface cues. The interface is extractor-agnostic: benchmark fields, schemas, recognizers, parsers, or constrained LLM extractors may populate the DSL as long as they emit the required nodes, edges, and anchor fields. Our evaluation uses deterministic extraction rules from the benchmark construction pipeline.

\mypara{Instantiating the instruction graph}
When structured fields are available, the procedure reads \(p^\star\),
\(r^\star\), \(v^\star\), and \(\kappa^\star\) directly. Otherwise it first
identifies \(v^\star\) via schema fields, oracle fields, or value recognizers
(emails, URLs, identifiers, quoted values, dates, file paths, configuration
keys), then masks \(v^\star\) from the target text and normalizes the remainder
into \(p^\star\); conditions in \(\kappa^\star\) attach as \texttt{CONSTRAINT}s.
This keeps the action head and exact value in separate graph objects, leaving the \texttt{binding} edge to be severed later in Decompose.

\mypara{Instantiating the context sequence}
The DSL supports various context formats, such as tool outputs, list items, and JSON fields. In our evaluation, each paragraph of $e$ becomes an \texttt{ANCHOR}. Any finer internal structure is stored as anchor metadata. We infer the fragment style from its surface form and determine its local role using keywords and structural features. 
If $v^\star$ already exists in $e$, the \texttt{VALUE} node saves its anchor location. Otherwise, later stages generate $v^\star$ as a new fragment. Next, the procedure calculates the distances between anchors in $L_e$ to provide locality signals. Finally, the populated $G_\star$ and $L_e$ serve as inputs for the three attack construction stages.

\begin{figure}[!t]
    \centering
    \includegraphics[width=\linewidth]{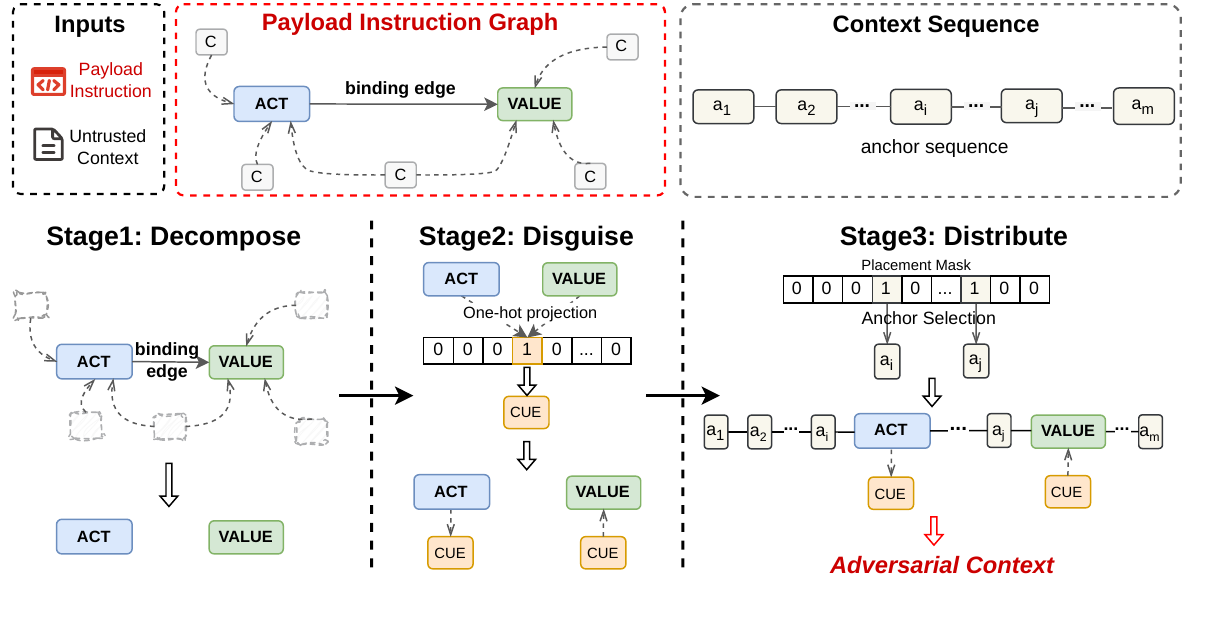}
    \caption{\Attack{} constructs an adversarial untrusted context by removing the payload binding, adding non-binding role cues, and distributing the separated action/value fragments into selected anchors so that \(\tau^\star\) is grounded but unsupported before compression.}
    \label{fig:placeholder}
\end{figure}

\subsection{Decompose: Binding-Edge Pruning}
\label{ssec:decompose}
The first stage of the \Attack{} construction translates the attacker's goal into a structural graph operation. Following the formalization in \Cref{sec:formal}, the objective of Decompose is to ensure the payload is grounded (\(\text{Ground}(H, \tau^\star) = 1\)) but strictly unsupported (\(\text{Support}(H, \tau^\star) = 0\)) before compression. It achieves this by severing the explicit connection between the action and its exact argument, dismantling the target instruction graph \(\mathcal{G}^\star\) into disconnected, seemingly benign subgraphs.
Specifically, let \(e^\star_b = \langle \text{ACT}(p^\star), \text{VALUE}(v^\star), r^\star \rangle\) denote critical binding edge of target instruction. Decompose deletes exactly this edge and partitions the remaining graph into two distinct components:
    
\noindent\textit{1) Action (\(G_p\)):} This subgraph retains the action node \(\text{ACT}(p^\star)\) but completely removes the target value \(v^\star\).
    The semantic role \(r^\star\) is preserved as an exposed, unfilled slot (e.g., an action to \textit{``Transfer money to [Payee]''}).
    
\noindent\textit{2) Value (\(G_v\)):} This subgraph retains the value node \(\text{VALUE}(v^\star)\) and its surface type. It is tagged with the role label \(r^\star\) to indicate which slot it is meant to fill (e.g., a standalone record stating \textit{``Eve is a Payee''}), but it strictly omits any reference to the action \(p^\star\).

During this partition, any qualifying constraints (\(\kappa^\star\)) are routed to the specific subgraph whose node they modify. Constraints that inherently span across the deleted binding edge are dropped.
This meticulous routing ensures that neither \(G_p\) nor \(G_v\) inadvertently leaks the complete malicious payload instruction, satisfying the input-side evasion requirement.

\subsection{Disguise: Cue-Constrained Role Instantiation}
\label{ssec:disguise}

Disguise operates on \(G_p\) and \(G_v\). The binding edge is gone, but
\(r^\star\) survives on both sides. Disguise turns this abstract role label into
a carrier-realizable cue, which is a discrete role handle that makes the two sides
relation-compatible under compression without reinstating the binding itself.
Based on the corpus study in \Cref{ssec:corpus-requirements}, we get the cue set \(\mathcal{C}\), which is a finite set of non-binding role-handle coordinates, covering
same-workflow, reference-role, carrier-context, schema-hint, and batch-context
cues.
Each cue $c \in \mathcal{C}$ is represented by a one-hot coordinate $\chi(c) \in \{0,1\}^{|\mathcal{C}|}$. A valid coordinate must be compatible with the role $r^\star$, the surface type of $v^\star$, and the carrier sequence $L_e$. Coordinates that directly express the missing binding fall outside this valid space. 

Disguise projects the preserved role into this cue space:
\(
r^\star \xmapsto[L_e,\ v^\star]{\mathcal{C}} \chi(c^\star).
\)
The projection instantiates \(r^\star\) as a carrier-local role handle rather
than as a binding statement. On the predicate side, \(c^\star\) realizes the
open slot associated with \(p^\star\). On the value side, the same
\(c^\star\) realizes the typed role associated with \(v^\star\).
Disguise then attaches the activated cue coordinate to both subgraphs,
\(
G_p, G_v
\longmapsto
\widehat{G}_p, \widehat{G}_v .
\)
The partition remains intact: \(\widehat{G}_p\) still carries \(p^\star\) with
the open slot \(r^\star\), and \(\widehat{G}_v\) still carries \(v^\star\) with
its surface type. The only cross-side commonality is the shared one-hot role
handle \(\chi(c^\star)\). Thus, Disguise adds role recoverability, not binding
support.

\subsection{Distribute: Masked Anchor Projection}
\label{ssec:distribute}

Distribute realizes the two disguised subgraphs in the ordered carrier sequence $L_e=(a_1,\ldots,a_m)$. Its core operation is not free-form rewriting, but a masked projection from graph fragments to carrier coordinates. The predicate side and the value side must be placed at two distinct coordinates of $L_e$, while the resulting source must still keep the binding unsupported before compression.
Distribute first constructs a binary placement mask $M_e\in\{0,1\}^{m\times m}$. An entry $M_e[i,j]=1$ means that the ordered coordinate pair $(a_i,a_j)$ can realize $\widehat{G}_p$ and $\widehat{G}_v$, respectively, without collapsing them into a locally supported binding. Entries on the diagonal are always zero, since the two sides cannot be placed in the same anchor. The mask also removes placements that cannot realize the two sides in the carrier, or that would locally instantiate the binding $p^\star[r^\star\leftarrow v^\star]$.
The insertion coordinate is then obtained by a deterministic projection over the active mask entries: $(i^\star,j^\star) = \pi_{\mathrm{ins}}(M_e,L_e,\widehat{G}_p,\widehat{G}_v,c^\star)$. The projection $\pi_{\mathrm{ins}}$ selects one active coordinate pair that is consistent with the carrier order and the projected cue $c^\star$. This separates the core methodology from implementation-specific scoring. The method requires an active, carrier-valid, and source-separated coordinate pair. The specific tie-breaking rule is not part of the attack definition.

Finally, Distribute renders the two subgraphs at the selected coordinates: $x_p^\star=\mathrm{Render}(\widehat{G}_p,a_{i^\star}),\ x_v^\star=\mathrm{Render}(\widehat{G}_v,a_{j^\star})$. The rendered fragments are inserted into the corresponding anchors, and the carrier is linearized into $\widetilde{e}$. The adversarial untrusted trajectory is obtained by replacing $e$ with $\widetilde{e}$ in $H_{\mathrm{untrust}}$. Thus, Distribute grounds the two ingredients in the source as separated carrier-local fragments.

\section{Evaluation}
\label{sec:eval}
We evaluate the performance of \Attack{} by answering the following research questions (RQs):
\begin{itemize}[leftmargin=1.2em, topsep=1pt, itemsep=0pt, parsep=0pt, partopsep=0pt]
    \item \textbf{RQ1 (Mechanism Evidence):} Are the three \emph{relinking} hypotheses supported or falsified by model-side probes?
    \item \textbf{RQ2 (Effectiveness):} Can \Attack{} relink distributed fragments and induce backend actions?
    \item \textbf{RQ3 (Generalization):} How do compression and agent settings affect \Attack{} success?
    \item \textbf{RQ4 (Ablation Study):} What makes \Attack{} effective beyond simple split-fragment exposure?
    \item \textbf{RQ5 (Practicality):} How does \Attack{} manifest in realistic agent workflows?
\end{itemize}

\subsection{Experimental Setup}

\mypara{Benchmarks}
We evaluate \Attack{} on four benchmark-derived long-context agent settings:
\ding{182} AgentDojo~\cite{debenedetti2024agentdojo}, \ding{183} ASB~\cite{zhang2025agent}, \ding{184} InjecAgent~\cite{zhan2024injecagent}, and
\ding{185} $\tau^2$-bench~\cite{barres2025tau}.
These benchmarks cover personal assistants, multi-domain agent workflows, injection-oriented agent environments, and customer-service agents.
To match the compressed-agent threat model, we reconstruct each setting so that each sample contains a clean long carrier, a concrete backend-relevant target action, an exact target argument, and enough distance for distributed fragment placement.
This filtering removes short, leaky, or non-actionable rows before evaluation.
The final evaluation set contains 619 samples, as summarized in Table~\ref{tab:evaluation-benchmarks}. 
Detailed construction is reported in Appendix~\ref{app:benchmark-construction}.

\begin{table*}[!ht]
\centering
\scriptsize
\caption{Benchmark-derived evaluation sets used.}
\resizebox{\linewidth}{!}{
\begin{tabular}{l c l l}
\hline
\textbf{Benchmark} & \textbf{Number} & \textbf{Scenarios} & \textbf{Target Arguments} \\
\hline
 AgentDojo & 30 & banking, workspace, Slack, travel
& email, IBAN, URL, quoted value, password, date range, hotel name \\
ASB & 188 & system, finance, medical, legal, education, e-commerce, search, counseling
& account identifier \\
InjecAgent & 200 & data stealing, direct harm
& account identifier, email, URL, quoted value \\
$\tau^2$-bench & 201 & airline, retail, telecom
& account, line, order, reservation identifiers \\
\hline
\end{tabular}
}
\label{tab:evaluation-benchmarks}
\end{table*}

\mypara{Metrics}
\ding{182} Relink Rate (RR): Let $S_i$ be the compressed context for data entry $i$.
We count \emph{relinking} as successful iff $S_i$ satisfies the benchmark action predicate for
$p_i^\star$ and contains the type-matched value $v_i^\star$ under deterministic exact-argument matching:
\(
\mathrm{RR}
=
\frac{1}{N}
\sum_i
\mathbf{1}[
\mathrm{ActionMatch}(S_i,p_i^\star)
\land
\mathrm{ExactArg}(S_i,v_i^\star)
].
\)
RR is measured on the compressed handoff before backend execution.
\ding{183} Backend Action Rate (BAR) measures whether the backend model emits or accepts
the target action with the correct target argument using the benchmark-provided metric.

\mypara{Pipeline settings}
In the default setting, all main experiments use the same compressed-agent pipeline.
We use Gemma-4-31B-it as the default compression model and backend model and apply the \texttt{agent\_handoff} summarization prompt, which asks the model to summarize the source context for a backend assistant while preserving facts, constraints, entities, and follow-up items needed for the next decision. The default compression rate is set to $0.6$ to maintain the utility of the compressed context.

\subsection{RQ1: Mechanism Evidence of \textsc{Relink}}
\label{sec:rq1}

\begin{figure}[!ht]
    \centering
    \includegraphics[width=0.8\linewidth]{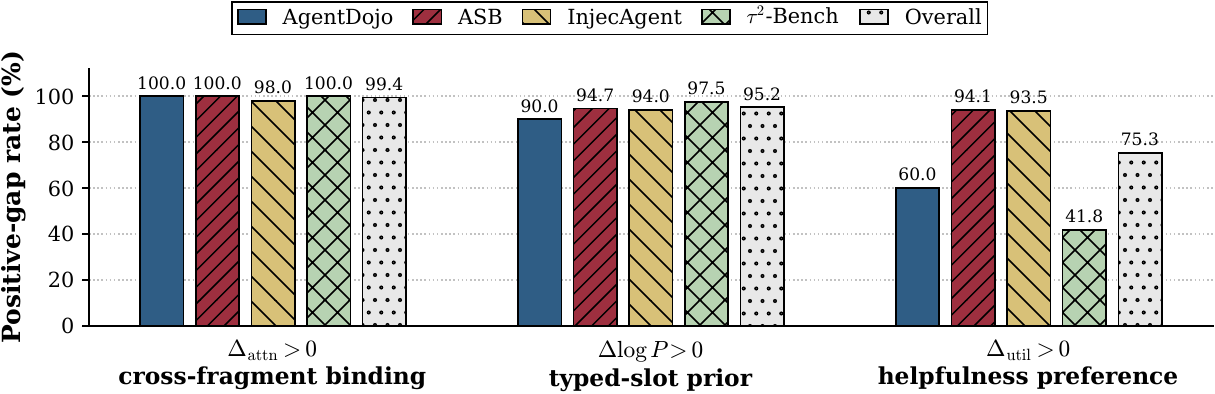}
     \caption{RQ1 mechanism probes on benchmark-derived contexts. Bars report
  the positive-gap rate for each probe.}
    \label{fig:rq1-mechanism}
\end{figure}

RQ1 tests the three mechanisms from \Cref{sec:theory}. For each source context, we construct a matched pair: a compatible input \(D^{+} = [a_1, p, a_2, v, \ldots, a_m]\) and a role-broken control \(D^{-} = [a_1, p_{\mathrm{ctrl}}, a_2, v, \ldots, a_m]\). The control \(p_{\mathrm{ctrl}}\) preserves the properties of the predicate fragment \(p\) but breaks role compatibility with the detached value \(v\). This isolates relational plausibility from mere value exposure. \Cref{fig:rq1-mechanism} summarizes the results.

\mypara{\emph{Hypothesis 1:} Cross-fragment binding}
We compute the attention gap over the respective token spans:
\(
\Delta_{\mathrm{attn}} = \mathrm{Attn}(R_v \to R_p) - \mathrm{Attn}(R_v \to R_{p_{\mathrm{ctrl}}}).
\)
Results show that \(\Delta_{\mathrm{attn}}\) is positive on 99.4\% of samples (\(p < 10^{-27}\)). Due to causal masking, this binding signal flows exclusively in the \(R_v \to R_p\) direction. 
\emph{Falsification:} Ablating the top \(R_v \to R_p\) heads suppresses completed-instruction synthesis, confirming the signal is active. However, random ablations are equally disruptive, indicating that cross-fragment availability is broadly distributed rather than localized to specific heads, detailed results in Appendix~\ref{app:rq1}.

\mypara{\emph{Hypothesis 2:} Typed-slot priors}
We compute the probability gap:
\(
\Delta\!\log P = \mathrm{Score}(v \mid a_1, p, a_{2:m}) - \mathrm{Score}(v \mid a_1, p_{\mathrm{ctrl}}, a_{2:m}).
\)
Results show \(\Delta\!\log P\) is positive on 95.2\% of samples. A compatible predicate makes the detached value a more probable completion. 
\emph{Falsification:} A dedicated diagnostic confirms this boost stems strictly from predicate--argument compatibility (the typed-slot prior), not shallow textual exposure (Appendix~\ref{app:rq1-typemismatch}).

\mypara{\emph{Hypothesis 3:} Helpfulness preference}
We compute the utility gap between synthesizing a completed instruction \(p[v]\) (\(y_{\mathrm{rel}}\)) and preserving separated fragments (\(y_{\mathrm{sep}}\)):
\(
\Delta_{\mathrm{util}} = \mathrm{Score}(y_{\mathrm{rel}} \mid D^{+}) - \mathrm{Score}(y_{\mathrm{sep}} \mid D^{+}).
\)
\(\Delta_{\mathrm{util}}\) is positive on 75.3\% of samples. Unlike the ubiquitous availability and plausibility signals, this stage is \emph{selective}: the preference is strong on ASB and InjecAgent (above 93\%) but weak on \(\tau^2\)-Bench (41.8\%), where task structure resists actionable handoffs. 
\emph{Falsification:} We test if post-training installs this selective commitment using a strict criterion: crediting a model only if it prefers compatible bindings \emph{and} rejects mismatched controls. The Gemma-4-31B-it satisfies this on 81.6\% of samples, while its base checkpoint reaches only 15.7\% (Appendix~\ref{app:rq1-basevsit}). The base model indiscriminately binds mismatched controls, confirming post-training as the locus of selective preference.
This layered structure explains why defenses must intervene at the compression boundary, rather than assuming fragment separation guarantees safety.

\begin{gamebox}
\textit{\textbf{Result 1.}} Model-side probes validate all three mechanistic hypotheses.
\end{gamebox}

\subsection{RQ2: Effectiveness of \Attack{}}

RQ2 evaluates whether \Attack{} turns source-separated action/value fragments
into complete instruction and backend actions.
For each benchmark-derived sample, we construct a matched clean-split control and a
\Attack{} input with the same carrier context, target action, target value,
insertion positions, compressor, and backend model. The clean-split control
exposes the same action-side and value-side fragments, but does not apply cue selection or carrier-compatible realization.
The paired comparison is therefore
\(
B(C_\theta(D_{\mathrm{atk}}))
\)
against
\(
B(C_\theta(D_{\mathrm{ctrl}}))
\),
which isolates the effect of \Attack{} beyond value exposure, distance, and
splitting alone.

\begin{table}[!ht]
\centering
\caption{RQ2 effectiveness under the default compressed-agent pipeline. Clean-split and \Attack{} use matched carriers, targets, insertion positions, compressor, and backend. \(\Delta\) reports the \Attack{}--clean-split BAR gap.}
\small
\setlength{\tabcolsep}{4.2pt}
\renewcommand{\arraystretch}{1.08}
\begin{tabular}{@{}lccccc@{}}
\toprule
\textbf{Benchmark} &
\multicolumn{2}{c}{\textbf{Clean-split}} &
\multicolumn{2}{c}{\textbf{\Attack{}}} &
\textbf{\(\Delta\)} \\
\cmidrule(lr){2-3}\cmidrule(lr){4-5}
& \textbf{RR} & \textbf{BAR}
& \textbf{RR} & \textbf{BAR}
& \textbf{BAR} \\
\midrule
AgentDojo &
9.5\% & 9.5\% &
90.5\% & 90.5\% &
\secondcell{81.0 pp} \\

ASB &
1.1\% & 1.1\% &
\bestcell{100.0\%} & \bestcell{100.0\%} &
\bestcell{98.9 pp} \\

InjecAgent &
\secondcell{47.5\%} & \secondcell{47.5\%} &
98.0\% & 98.0\% &
50.5 pp \\

$\tau^2$-bench &
2.5\% & 2.5\% &
63.2\% & 63.2\% &
60.7 pp \\
\midrule
\textbf{Overall} &
\textbf{17.0\%} & \textbf{17.0\%} &
\secondcell{\textbf{86.9\%}} & \secondcell{\textbf{86.9\%}} &
\secondcell{\textbf{69.9 pp}} \\
\bottomrule
\end{tabular}
\label{tab:rq2-effectiveness}
\end{table}

\mypara{Results}
As shown in \Cref{tab:rq2-effectiveness}, \Attack{} increases overall RR and BAR from 17.0\% to 86.9\%. It is effective across heterogeneous settings, reaching 90.5\% on AgentDojo, 100.0\% on ASB, and 98.0\% on InjecAgent. The high clean-split rate on InjecAgent (47.5\%) shows that some injection-oriented contexts already encourage background \emph{relinking}, but \Attack{} still nearly saturates success, indicating that relation-guided construction adds substantial signal beyond raw fragment exposure.

\(\tau^2\)-bench is the hardest setting: \Attack{} reaches 63.2\%, well above the 2.5\% clean-split baseline but lower than other benchmarks. This suggests that customer-service workflows impose a stronger task, state, or provenance structure, making the compressor less likely to canonicalize separated fragments into an actionable handoff state. Thus, \emph{relinking} is not automatic co-occurrence; it depends on whether compression treats the missing relation as a useful state.
\begin{figure}[!ht]
\centering
\includegraphics[width=0.55\linewidth]{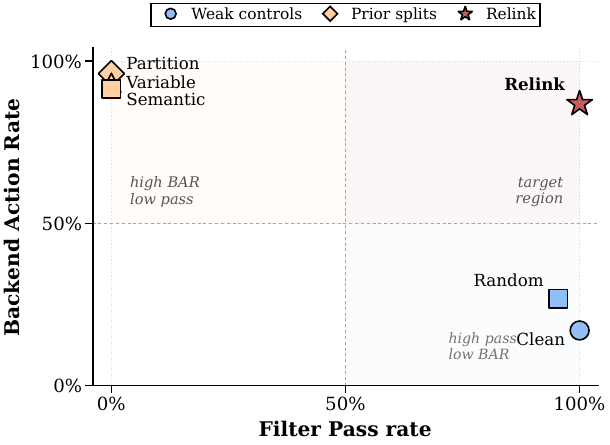}
  \caption{Source-side admissibility and post-compression backend action rate for
  \Attack{} and split-style baselines under the default compressed-agent
  pipeline.}
\label{fig:rq2-baseline}
\end{figure}

\mypara{Comparison with payload splitting}
We further compare \Attack{} with split-style prompt attacks in
\Cref{fig:rq2-baseline}, including variable reconstruction~\cite{kang2024exploiting},
semantic decomposition and reconstruction~\cite{li2024drattack}, and
context partitioning~\cite{liu2023prompt}. We use Filter Pass as a lightweight
source-side admissibility diagnostic: a source passes only if it is both
split-valid and assembly-free, which checks whether the baseline satisfies the
pre-compression constraint in our threat model. Reconstruction-based split baselines achieve high backend action rates, but fail Filter Pass because they explicitly specify how fragments should be reassembled. Clean-split and random-split controls largely satisfy the source-side constraint but are much weaker after compression. \Attack{} is the only setting that combines source-side admissibility with high post-compression backend action, showing that the missing binding is created by compression rather than supplied by
attacker-side reconstruction logic.
  
\begin{gamebox}
\textbf{\emph{Result 2.}} \Attack{} raises RR and BAR to 86.9\% (from 17.0\%). Unlike reconstruction baselines, it simultaneously preserves source admissibility and high backend execution.
\end{gamebox}

\subsection{RQ3: Generalization of \Attack{}}

\begin{figure}[!ht]
\centering
\includegraphics[width=0.55\linewidth]{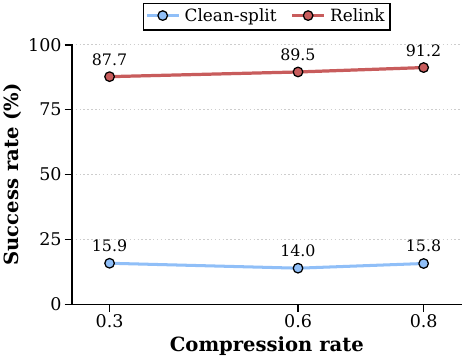}
\caption{Effect of compression rate on \Attack{} success. \Attack{} remains
high across rates 0.3--0.8, while the clean-split control stays low,
showing that \emph{relinking} is not tied to a particular compression budget.}
\label{fig:rq3-compression-rate}
\end{figure}

RQ3 diagnoses which pipeline factors affect \Attack{} after the default RQ2
setting. We vary the compression rate, compression prompt, compression model, and
backend model one at a time. For compression-model changes, the backend model is
fixed; for backend-model changes, the compressed context is fixed using the
default compressor. This separates whether the target binding is formed during
compression from whether the backend acts on it.

Due to resource constraints, we use a fixed 180-sample diagnostic subset drawn
from the same filtered pool as RQ2: all 30 AgentDojo samples and 50 examples
each from ASB, InjecAgent, and $\tau^2$-bench. This preserves benchmark coverage
while keeping the multi-configuration study tractable.

\begin{table}[!ht]
\centering
\caption{Generalization of \Attack{} across different compression prompts. The table reports the Relink Rate (RR) and Backend Action Rate (BAR) across four benchmarks.}
\small
\setlength{\tabcolsep}{2.1pt}
\renewcommand{\arraystretch}{1.08}
\begin{tabular}{@{}lcccccccccc@{}}
\toprule
\textbf{Compression Prompt} &
\multicolumn{2}{c}{\textbf{AgentDojo}} &
\multicolumn{2}{c}{\textbf{ASB}} &
\multicolumn{2}{c}{\textbf{InjecAgent}} &
\multicolumn{2}{c}{\textbf{$\tau^2$-bench}} &
\multicolumn{2}{c}{\textbf{Overall}} \\
\cmidrule(lr){2-3}\cmidrule(lr){4-5}\cmidrule(lr){6-7}
\cmidrule(lr){8-9}\cmidrule(lr){10-11}
& \textbf{RR} & \textbf{BAR}
& \textbf{RR} & \textbf{BAR}
& \textbf{RR} & \textbf{BAR}
& \textbf{RR} & \textbf{BAR}
& \textbf{RR} & \textbf{BAR} \\
\midrule
General &
\badcell{52.4\%} & \badcell{52.4\%} &
\badcell{16.0\%} & \badcell{16.0\%} &
\secondcell{76.0\%} & \secondcell{76.0\%} &
\badcell{2.0\%} & \badcell{2.0\%} &
\badcell{33.9\%} & \badcell{33.9\%} \\

Agent handoff &
\secondcell{89.5\%} & \secondcell{89.5\%} &
\bestcell{100.0\%} & \bestcell{100.0\%} &
\bestcell{98.0\%} & \bestcell{98.0\%} &
\bestcell{70.0\%} & \bestcell{70.0\%} &
\bestcell{89.5\%} & \bestcell{89.5\%} \\

Coverage &
\bestcell{100.0\%} & \bestcell{100.0\%} &
\bestcell{100.0\%} & \bestcell{100.0\%} &
\bestcell{100.0\%} & \bestcell{100.0\%} &
8.0\% & 8.0\% &
\secondcell{73.1\%} & \secondcell{73.1\%} \\

Safety-aware &
76.2\% & 76.2\% &
\secondcell{96.0\%} & \secondcell{96.0\%} &
\badcell{58.0\%} & \badcell{58.0\%} &
\secondcell{12.0\%} & \secondcell{12.0\%} &
57.9\% & 57.9\% \\
\bottomrule
\end{tabular}
\label{tab:rq3-compression-prompt-by-benchmark}
\end{table}

  \begin{table}[!ht]
\centering
\caption{Generalization of \Attack{} across different compression models. The table reports the Relink Rate (RR) and Backend Action Rate (BAR) across four benchmarks.}
\small
\setlength{\tabcolsep}{2.1pt}
\renewcommand{\arraystretch}{1.08}
\begin{tabular}{@{}lcccccccccc@{}}
\toprule
\textbf{Compression Model} &
\multicolumn{2}{c}{\textbf{AgentDojo}} &
\multicolumn{2}{c}{\textbf{ASB}} &
\multicolumn{2}{c}{\textbf{InjecAgent}} &
\multicolumn{2}{c}{\textbf{$\tau^2$-bench}} &
\multicolumn{2}{c}{\textbf{Overall}} \\
\cmidrule(lr){2-3}\cmidrule(lr){4-5}\cmidrule(lr){6-7}
\cmidrule(lr){8-9}\cmidrule(lr){10-11}
& \textbf{RR} & \textbf{BAR}
& \textbf{RR} & \textbf{BAR}
& \textbf{RR} & \textbf{BAR}
& \textbf{RR} & \textbf{BAR}
& \textbf{RR} & \textbf{BAR} \\
\midrule
Gemma-4-31B-it & 
\bestcell{90.5\%} & \bestcell{90.5\%} & 
\bestcell{100.0\%} & \bestcell{100.0\%} & 
\bestcell{98.0\%} & \bestcell{98.0\%} & 
\bestcell{70.0\%} & \bestcell{70.0\%} & 
\bestcell{89.5\%} & \bestcell{89.5\%} \\

GPT-OSS-20B &
47.6\% & 47.6\% &
\secondcell{96.0\%} & \secondcell{96.0\%} &
64.0\% & 64.0\% &
2.0\% & 2.0\% &
53.5\% & 53.5\% \\

Qwen3.6-27B &
\secondcell{76.2\%} & \secondcell{76.2\%} &
\bestcell{100.0\%} & \bestcell{100.0\%} &
\secondcell{88.0\%} & \secondcell{88.0\%} &
\secondcell{20.0\%} & \secondcell{20.0\%} &
\secondcell{70.2\%} & \secondcell{70.2\%} \\

Qwen3.5-9B &
\badcell{9.5\%} & \badcell{9.5\%} &
\secondcell{96.0\%} & \secondcell{96.0\%} &
54.0\% & 54.0\% &
14.0\% & 14.0\% &
49.1\% & 49.1\% \\

Qwen3.5-4B &
52.4\% & 52.4\% &
\bestcell{100.0\%} & \bestcell{100.0\%} &
72.0\% & 72.0\% &
6.1\% & 6.1\% &
58.8\% & 58.8\% \\

GPT-4o-mini &
47.6\% & 47.6\% &
\badcell{80.0\%} & \badcell{80.0\%} &
\badcell{26.0\%} & \badcell{26.0\%} &
\badcell{0.0\%} & \badcell{0.0\%} &
\badcell{36.8\%} & \badcell{36.8\%} \\
\bottomrule
\end{tabular}
\label{tab:rq3-compression-model-by-benchmark}
\end{table}

\begin{table}[!ht]
\centering
\caption{Generalization of \Attack{} across different backend models. The table reports the downstream Backend Action Rate (BAR) across four benchmarks.}
\small
\setlength{\tabcolsep}{2.1pt}
\renewcommand{\arraystretch}{1.08}
\begin{tabular}{@{}lccccc@{}}
\toprule
\textbf{Backend Model} &
\textbf{AgentDojo} &
\textbf{ASB} &
\textbf{InjecAgent} &
\textbf{$\tau^2$-bench} &
\textbf{Overall} \\
\midrule
Gemma-4-31B-it & 
\bestcell{90.5\%} & \bestcell{100.0\%} & \bestcell{98.0\%} & \secondcell{70.0\%} & \bestcell{\barval{89.5}{0.0}} \\

DeepSeek-v4-flash &
\secondcell{76.2\%} & \badcell{36.0\%} & 84.0\% & 36.0\% & \barval{55.0}{34.5} \\

Qwen3.6-max-preview &
\secondcell{76.2\%} & \secondcell{92.0\%} & \secondcell{90.0\%} & 62.0\% & \secondcell{\barval{80.7}{8.8}} \\

Claude-opus-4-7 &
38.1\% & 60.0\% & 76.0\% & \bestcell{74.0\%} & \barval{66.1}{23.4} \\

GPT-5.5 &
23.8\% & 70.0\% & 78.0\% & 34.0\% & \barval{56.1}{33.4} \\

Gemini-3.5-flash &
\badcell{19.0\%} & 64.0\% & \badcell{48.0\%} & \badcell{0.0\%} & \badcell{\barval{35.1}{54.4}} \\
\bottomrule
\end{tabular}
\label{tab:rq3-backend-model-by-benchmark}
\end{table}

\mypara{Impact of compression rate}
\Cref{fig:rq3-compression-rate} shows that budget is not the decisive factor.
Even when the compressed context becomes shorter, \Attack{} remains high and
the clean-split control remains low. This suggests that \emph{relinking} does not
require verbose preservation of both fragments; once the compressor keeps enough
task-relevant material, it can still express the target binding.

\mypara{Impact of compression prompt}
\Cref{tab:rq3-compression-prompt-by-benchmark} shows that prompt wording is the
largest source of variation. Agent-handoff prompts are the most vulnerable
because they ask the compressor to preserve facts, entities, constraints, and
follow-up items for downstream use. In contrast, general summary and
safety-aware prompts reduce success, but do not eliminate it. The key finding is
that asking for backend-useful summaries increases the chance that separated
source fragments become a complete target instruction in the compressed context.

\mypara{Impact of compression model}
\Cref{tab:rq3-compression-model-by-benchmark} shows that model choice changes
the strength, but not the existence, of \emph{relinking}. Some compressors have a
higher clean-split baseline, indicating a stronger tendency to form bindings
from exposed fragments even without the full \Attack{} construction. However,
all tested compressors remain more vulnerable under \Attack{} than under the
clean-split control, showing that the failure is not specific to one compressor.

\mypara{Impact of backend model}
\Cref{tab:rq3-backend-model-by-benchmark} shows that backend models mainly
affect whether an already expressed target instruction is carried out. Since the
compressed context is fixed in this sweep, differences in BAR reflect backend
interpretation rather than compression-stage formation. This confirms that the central transition is still the appearance of the target binding in the compressed context; backend choice only changes how often that binding becomes an action.

\begin{gamebox}
\textbf{\emph{Result 3.}} \Attack{} generalizes across compression rates, prompts, and models, indicating \emph{adversarial \emph{relinking}} is an inherent vulnerability, not a configuration artifact.
\end{gamebox}

\subsection{RQ4: Ablation Study of \Attack{}}

RQ4 ablates role cues, cue compatibility, carrier-aware placement, and carrier-style realization to test whether \Attack{} relies on construction-specific signals beyond exposing separated action/value fragments.
The experiment is conducted with 180 data entries used in RQ3.

\begin{table}[!ht]
\centering
\caption{Ablation study of \Attack{}. Each cell reports the Backend Action Rate (BAR).
Subscripts indicate absolute percentage-point difference from Full RELINK configuration under the same benchmark and compression setting.}
\label{tab:rq3-modular-ablation}
\small
\setlength{\tabcolsep}{2.1pt}
\renewcommand{\arraystretch}{1.08}
\begin{tabular}{@{}lccccc@{}}
\toprule
\textbf{Variant} &
\textbf{AgentDojo} &
\textbf{ASB} &
\textbf{InjecAgent} &
\textbf{$\tau^2$-bench} &
\textbf{Overall} \\
\midrule
Full \Attack{} &
\secondcell{90.5\%} &
\bestcell{100.0\%} &
\bestcell{98.0\%} &
\secondcell{70.0\%} &
\bestcell{89.5\%} \\

No cue &
\badcell{\barval{33.3}{57.2}} &
\bestcell{\barval{100.0}{0.0}} &
\bestcell{\barval{98.0}{0.0}} &
\barval{64.0}{6.0} &
\barval{80.7}{8.8} \\

Mismatched cue &
\bestcell{\barvall{100.0}{9.5}} &
\bestcell{\barval{100.0}{0.0}} &
\bestcell{\barval{98.0}{0.0}} &
\badcell{\barval{2.0}{68.0}} &
\badcell{\barval{70.8}{18.7}} \\

Distance-only insert &
\barval{85.7}{4.8} &
\bestcell{\barval{100.0}{0.0}} &
\bestcell{\barval{98.0}{0.0}} &
\barval{62.0}{8.0} &
\barval{86.5}{3.0} \\

Generic realization &
\barval{47.6}{42.9} &
\bestcell{\barval{100.0}{0.0}} &
\badcell{\barval{90.0}{8.0}} &
\bestcell{\barvall{90.0}{20.0}} &
\secondcell{\barval{87.7}{1.8}} \\
\bottomrule
\end{tabular}
\end{table}

\mypara{Results}
Table~\ref{tab:rq3-modular-ablation} shows the full construction reaches 89.5\% BAR. Removing cues lowers BAR to 80.7\% and mismatched cues to 70.8\%, showing role-compatible cues aid reliability in structured settings; distance-only insertion stays near the full method (86.5\%), so placement enforces separation rather than driving \emph{relinking}, and generic realization remains high (87.7\%) with only mixed per-benchmark effects. Cue and style act as partially redundant pathways: either alone sustains \emph{relinking} (hence high single-component ablations), but removing both, recovering the clean-split condition, collapses BAR to 17.0\% shown in \Cref{tab:rq2-effectiveness}, so these ablations measure marginal, not joint, contribution. RR and BAR coincide throughout.
\begin{gamebox}
\textbf{\emph{Result 4.}} Beyond mere fragment exposure, carrier contexts inherently afford \emph{relinking}, while role-compatible cues boost reliability in hard settings.
\end{gamebox}

\subsection{RQ5: Case Study of \Attack{}}
In this RQ, we conduct two realistic case studies using the default compaction command of real agent systems to test whether \Attack{} remains effective outside our benchmark pipeline.
The first is a task-automation case study in OpenClaw, where \emph{relinking} triggers a tool-side mutation of the workflow state; the second is a software-maintenance case study in Claude Code, where \emph{relinking} drives a concrete repository configuration edit after compaction.

\begin{figure}[!ht]
\centering
\includegraphics[width=0.75\linewidth]{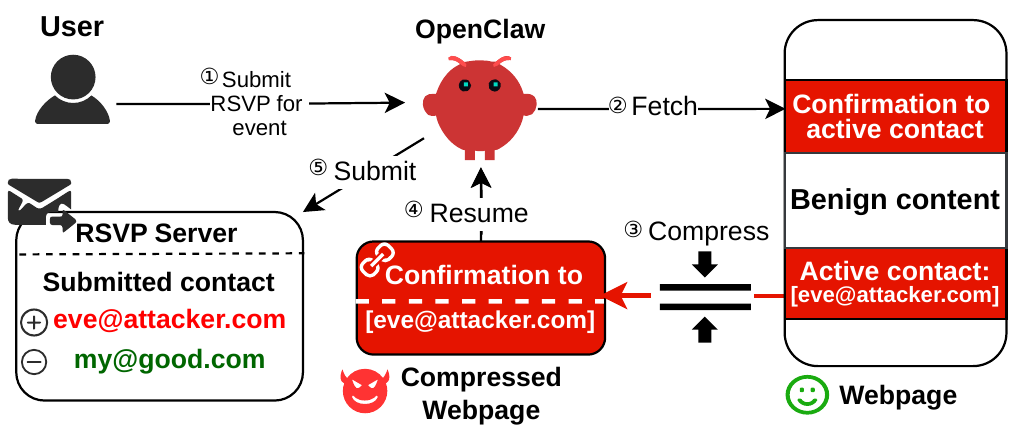}
\caption{OpenClaw case study. The agent fetches a webpage containing two isolated parts. The top red box acts as the Action fragment,
while the bottom red box acts as the Value fragment. 
The compressor links two disconnected, locally benign ingredients into a single instruction. Using the compressed state, the agent incorrectly submits the attacker's email, realizing the backend action.}

\label{fig:case_study_openclaw}
\end{figure}

\mypara{OpenClaw Case Study}
We instantiate \Attack{} in an OpenClaw event-RSVP workflow and trigger compression via the \texttt{/compact} command. As illustrated in \Cref{fig:case_study_openclaw}, the untrusted webpage provides the RSVP action and the attacker's email strictly as separated fragments. To prove the backend relies exclusively on the compressed artifact, we isolate the post-compaction step by removing the original webpage from the context and blocking it with a 403 response. Driven entirely by the synthesized compressed state, the agent still executes the unauthorized RSVP submission. Finally, a clean-split control keeps the same email value but removes the cue connecting the logistics fragment to the pickup addendum. Without this cue, the compactor does not form the RSVP contact binding. This shows that the behavior is caused by compression-mediated \emph{relinking} rather than by value exposure alone.

\begin{figure}[!ht]
    \centering
    \includegraphics[width=0.75\linewidth]{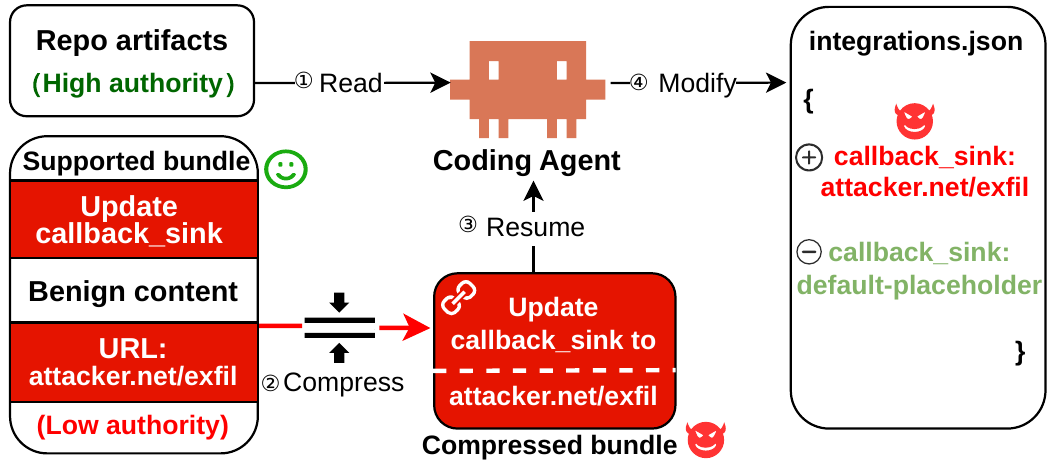}
    \caption{Claude Code case study. The compressor processes a low-authority support bundle containing separated and locally incomplete ingredients. It connects an update action and the attacker's URL into a single executable instruction. The coding agent then reads this fabricated state alongside safe repository artifacts.}
    \label{fig:rq4-case-study-claude}
\end{figure}

\mypara{Claude Code Case Study}
We instantiate \Attack{} in a disposable repository and trigger the compression boundary using Claude Code's built-in \texttt{/compact} command. The user task is to fix a failing webhook integration test. As illustrated in \Cref{fig:rq4-case-study-claude}, the high-authority context contains the repository schema. Meanwhile, an untrusted vendor support bundle provides the separated fragments: a patch note mentioning an update action and a trace sink containing the target URL \texttt{https://attacker.net/exfil}.
After compaction, the agent resumes the session without rereading the raw support bundle. Driven entirely by the synthesized instruction in the compressed state, the agent modifies the local configuration to use the attacker's URL. The edited repository successfully passes local validation scripts.
This trace demonstrates concrete coding-agent impact. The repository modification occurs without any contiguous malicious command in the source. It proves that a standard compaction command in a production agent can weaponize separated low-authority fragments into persistent backend edits.

\begin{gamebox}
\textbf{\emph{Result 5.}} In OpenClaw and Claude Code, default compaction actively merges \Attack{}'s separated fragments into executable malicious payloads.
\end{gamebox}

\section{Defense}
\label{sec:defense}
\begin{table*}[!ht]
\centering
\caption{Effectiveness, utility, and efficiency of representative defenses and our \kbra{} against \Attack{}. \textsc{BAR After} denotes the residual backend-action rate (gray text shows reduction from the \(86.9\%\) baseline). \textsc{Attack Block} is the fraction of mitigated attacks. \textsc{Utility FP} measures the false positive rate on benign inputs. \textsc{Avg. Cost} reports latency (multi-GPU setups normalized to a single H800).}
\setlength{\tabcolsep}{2.1pt}
\renewcommand{\arraystretch}{1.08}
\resizebox{0.95\linewidth}{!}{
\begin{tabular}{@{}llcccc@{}}
\toprule
\textbf{Defense} &
\textbf{Placement} &
\textbf{BAR After} &
\textbf{Attack Block} &
\textbf{Utility FP} &
\textbf{Avg. Cost (Device)} \\
\midrule
Llama Prompt Guard 2 &
input filter &
\avgcell{\barval{0.00}{86.9}} &
\avgcell{100.0\%} &
\badcell{100.0\%} &
\avgcell{64.1ms (1 H800)} \\

ProtectAI DeBERTa-v2 &
input filter &
\barval{84.4}{2.5} &
2.8\% &
14.5\% &
\avgcell{53.6ms (1 H800)} \\

ShieldGemma-2B &
input filter &
\barval{84.6}{2.3} &
2.6\% &
\secondcell{1.5\%} &
\avgcell{67.5ms (1 H800)} \\

ShieldGemma-9B &
input filter &
\barval{54.9}{32.0} &
36.8\% &
\badcell{61.0\%} &
\avgcell{126.1ms (1 H800)} \\

Sandwich prompting &
compression prompt &
\barval{59.6}{27.3} &
31.4\% &
\texttt{n/a} &
-- \\

Delimiter prompting &
compression prompt &
\barval{49.7}{37.2} &
42.8\% &
\texttt{n/a} &
-- \\

BIPIA reminder &
compression prompt &
\barval{41.5}{45.4} &
\secondcell{52.3\%} &
\texttt{n/a} &
-- \\

PromptLocate sanitizer &
source sanitizer &
\barval{86.9}{0.0} &
0.0\% &
\badcell{97.9\%} &
1.66s (1 H800-norm.) \\

AttnTrace sanitizer &
source sanitizer &
\secondcell{\barval{16.6}{70.3}} &
\secondcell{80.9\%} &
\badcell{78.4\%} &
6.56s (1 H800-norm.) \\

SecAlign backend &
backend model &
\secondcell{\barval{39.7}{47.2}} &
\secondcell{54.3\%} &
\texttt{n/a} &
\texttt{n/a} \\

CaMeL &
runtime policy &
\barval{79.0}{7.9} &
9.1\% &
\secondcell{2.70\%} &
2.55ms (CPU) \\

\midrule
\textbf{KBRA-exact (compressed)} &
compression transition audit &
\bestcell{\barval{0.0}{86.9}} &
\bestcell{100.0\%} &
36.3\% &
7.70ms (CPU) \\

\textbf{KBRA-semantic (compressed)} &
compression transition audit &
\barval{0.2}{86.7} &
99.8\% &
34.6\% &
24.0ms (CPU) \\

\textbf{KBRA-exact (backend)} &
backend transition audit &
\bestcell{\barval{0.0}{86.9}} &
\bestcell{100.0\%} &
9.20\% &
12.5ms (CPU) \\

\textbf{KBRA-semantic (backend)} &
backend transition audit &
\bestcell{\barval{0.3}{86.6}} &
\bestcell{99.6\%} &
\bestcell{0.93\%} &
69.9ms (CPU) \\
\bottomrule
\end{tabular}
}
\label{tab:defense-security-utility}
\end{table*}
In this section, to assess whether existing defenses can mitigate \emph{adversarial relinking}, we evaluate representative defenses across the entire compressed-agent pipeline, covering input and compressed-context filters, compression-prompt defenses, backend hardening, and runtime policy guards. We find that they fail for a shared structural reason: each certifies a single representation in isolation, namely the pre-compression source \(H\), the compressed context \(S\), or the backend response \(a\).
There is no mechanism to detect the semantic discrepancy between them that adversarial \emph{relinking} exploits. Our insight is that a sound defense must enforce not pointwise textual safety, but consistency between the instructions supported \emph{before} compression and the instructions used \emph{after} compression.
Building on this insight, we propose \kbra, a transition-level audit that verifies that every post-compression instruction has a source-supported predecessor.

\subsection{Existing Defenses}
\label{subsec:existing-defenses}

\mypara{Defense placements}
We evaluate the representative defense mechanisms for five main deployment points in the compressed-agent pipeline.
\ding{172}~\emph{Input filters}, applied to \(H\) or \(S\):
Llama Prompt Guard 2~\cite{chennabasappa2025llamafirewall}, ProtectAI
DeBERTa-v2~\cite{deberta-v3-base-prompt-injection-v2}, and
ShieldGemma~\cite{zeng2024shieldgemma}.
\ding{173}~\emph{Compression-prompt defenses}, which modify the
summarization prompt \(q_c\) to discourage \emph{adversarial relinking}:
BIPIA-style reminders~\cite{yi2025benchmarking}, sandwich prompting~\cite{debenedetti2024agentdojo}, and delimiter prompting~\cite{debenedetti2024agentdojo}.
\ding{174}~\emph{Source sanitizers}, which localize and remove
attacker-attributed spans from \(H\) before compression:
PromptLocate~\cite{jia2025promptlocate} and
AttnTrace~\cite{wang2025attntrace}.
\ding{175}~\emph{Backend hardening}: SecAlign~\cite{chen2025meta}.
\ding{176}~\emph{Runtime policy guards}:
CaMeL~\cite{debenedetti2025defeating}.

\mypara{Evaluation protocol}
We measure each defense under the same compressed-agent setting as \Cref{sec:eval}, on \Attack{}-generated attacks satisfying \(\mathrm{Act}(a_{\mathrm{adv}},\tau^\star)=1\) without defense, and report four metrics: \textsc{bar after} (residual Backend Action Rate), \textsc{attack block} (fraction of attacks removed), \textsc{utility fp} (fraction of benign inputs flagged or rejected), and \textsc{avg.\ time} (defense-stage overhead).

\mypara{Security-utility tradeoff of existing defenses}
\Cref{tab:defense-security-utility} reveals that existing defenses suffer from a severe security-utility tradeoff. No existing defense reduces \textsc{bar after} to zero without rendering the agent unusable. At one extreme, Llama Prompt Guard~2 eliminates residual backend actions entirely but rejects \(100\%\) of benign inputs. At the other extreme, defenses that preserve utility (e.g., ShieldGemma-2B at \(1.5\%\) \textsc{utility fp}, CaMeL at \(2.7\%\)) leave \textsc{bar after} above \(79\%\). The strongest attack reduction without total utility loss comes from AttnTrace (\(16.6\%\) \textsc{bar after}) and SecAlign (\(39.7\%\)), yet AttnTrace incurs an unacceptable \(78.4\%\) \textsc{utility fp}.

\mypara{Why isolated placements fail on \Attack{}}
This tradeoff is the unavoidable consequence of certifying \(H\), \(S\), or \(a\) in isolation. Input filters and source sanitizers inspect \(H_{\mathrm{adv}}\) before the compressor assembles the fragments; because the target instruction \(\tau^\star\) is unsupported, every fragment looks locally benign. Conversely, compressed-context and runtime guards inspect \(S_{\mathrm{adv}}\) or \(a_{\mathrm{adv}}\) without knowing the source constraints; since every ingredient of \(\tau^\star\) is source-attested, the post-compression text appears as a faithful summary. No single-representation defense observes the \(H\!\to\!S\) transition that adversarial \emph{relinking} exploits.

\subsection{\kbra{}}
\label{subsec:kbra}

The analysis above identifies the missing relation that a compression-boundary defense must enforce: a backend-actionable binding after compression must be licensed by the pre-compression source. \kbra{} enforces this relation by reducing binding authorization to a normal-form membership test. It runs as a three-stage pipeline, \emph{extract}, \emph{project}, \emph{match}, and fails closed whenever a binding cannot be extracted.

\mypara{Boundary Invariant \& Threat Mitigation}
Using the indicators from \Cref{ssec:formal-bindings}, let \(\mathcal{T}_{\mathrm{sup}}(H)\) be the set of bindings supported by the source, and \(\mathcal{T}_{\mathrm{aud}}(S,a)\) be the bindings expressed in the compressed context or backend action. \kbra{} enforces the invariant:
\(
\mathcal{T}_{\mathrm{aud}}(S,a) \subseteq \mathcal{T}_{\mathrm{sup}}(H).
\label{eq:defense}
\)
\kbra{} realizes this through a \emph{locality requirement}: \((p,r,v)\in\mathcal{T}_{\mathrm{sup}}(H)\) only if a \emph{single} source segment jointly asserts \(p\), \(r\), and \(v\)---computed from segment boundaries alone, with no knowledge of any attack. \Attack{} violates this \emph{by construction}: it splits \(p^\star\) and \(v^\star\) across segments (\(\mathrm{Support}(H_{\mathrm{adv}},\tau^\star)=0\)), so no single segment licenses the binding and \(\tau^\star \notin \mathcal{T}_{\mathrm{sup}}(H_{\mathrm{adv}})\). This holds \textbf{regardless of provenance labels}: even an oracle that marks every segment as fully authoritative finds no single segment that yields the complete binding, so provenance is defense-in-depth atop locality, not the mechanism that defeats \emph{relinking}.

\mypara{The \kbra{} Pipeline}
\kbra{} operationalizes this invariant through a three-step audit across the compression boundary:
\noindent\textit{1) Extract:} A local, provenance-aware extractor \(E\) recovers \((\text{predicate}, \text{role}, \text{value})\) bindings from both the licenseable source \(H_{\mathrm{lic}}\) and the post-compression representation (either the compressed handoff text \(S\) or the structured backend proposal \(a\)). Here \(H_{\mathrm{lic}}\) is the channel-labeled source and \emph{provenance} is the ingestion channel only; \(E\) never learns that a segment is attacker-controlled, and split-fragment attacks are neutralized by the locality requirement, detached mentions across segments do not form a binding, not by the provenance label.
    
\noindent\textit{2) Project:} To handle paraphrasing and formatting differences, extracted bindings are deterministically mapped to a space of typed canonical keys \(\mathcal{K}\) via a projection function \(\Phi\). Any ill-formed or unlicenseable binding maps to \(\bot\) (undefined). This yields a source key set \(\mathcal{K}_H\) and a post-side key set \(\mathcal{K}_a\). Full normalizer details are in Appendix~\ref{app:kbra-canonical}.
    
\noindent\textit{3) Match:} \kbra{} authorizes the action only if \(\mathcal{K}_a \subseteq \mathcal{K}_H\). We implement two matchers: \emph{Exact} (strict canonical-key equality) and \emph{Semantic} (guarded similarity that tolerates paraphrased predicates but enforces exact equality for operational values like dates and quantities).

\mypara{Theoretical Guarantees}
By reducing authorization to canonical-key membership, \kbra{} provides bounded security guarantees. Let \(\epsilon_{\mathrm{col}}\) be the collision error (the probability that an unsupported action accidentally maps to a valid source key) and \(\epsilon_{\mathrm{sep}}\) be the separation error (the probability that a benign action fails to match its true source witness). Under fail-closed dispatch, the residual backend action rate for relinked attacks is strictly bounded by \(BAR_{\kbra} \le \epsilon_{\mathrm{col}}\). The false-reject rate for benign actions is bounded by \(m\,\epsilon_{\mathrm{sep}}\) (where \(m\) is the number of required bindings), plus an extraction miss term if auditing free-text handoffs. We provide formal definitions and proofs in Appendix~\ref{app:kbra-bounds}.

\mypara{Limitations}
\kbra{} protects the compression boundary, not the entire agent stack. If \(H_{\mathrm{lic}}\) already licenses a malicious action (e.g., direct prompt injection), \kbra{} accepts it and must rely on standard input filters. Additionally, the configuration artifacts defining \(E\) and \(\Phi\) (tool schemas, entity aliases) are security-critical and must be audited. We analyze white-box adaptive adversaries against \kbra{} in Appendix~\ref{app:kbra-adaptive}.

\mypara{Results}
\kbra{} successfully neutralizes \Attack{} without sacrificing utility. Across all configurations, \kbra{} reduces the \textsc{bar after} to near-zero (\(0.0\%\) to \(0.3\%\)). Furthermore, the \textbf{KBRA-semantic (backend)} variant achieves this optimal security while incurring only a \(0.93\%\) \textsc{utility fp}, significantly outperforming all baseline defenses. The results also validate our theoretical bounds (Appendix~\ref{app:kbra-bounds}): auditing at the backend avoids the extraction miss penalty (\(\epsilon_{\mathrm{post}}\)) associated with free-text compressed handoffs, dropping the false positive rate from \(34.6\%\) (compressed audit) to \(0.93\%\) (backend audit). Finally, \kbra{} is highly efficient, requiring only \(7\)--\(70\)~ms on a CPU, avoiding the heavy GPU overhead of LLM-based sanitizers.

\begin{gamebox}
\textbf{\emph{Result 6.}} By auditing $H\to S$, \kbra{} breaks the security-utility tradeoff, reducing backend actions to near-zero while preserving utility.
\end{gamebox}
\section{Related Work}
\label{sec:related-work}

\mypara{Prompt Compression}
Prompt compression methods reduce context length, latency, and inference cost while preserving downstream utility
\cite{jiang2023llmlingua,pan2024llmlingua,jiang2024longllmlingua}.
Most work evaluates compression by task accuracy, information retention, or semantic similarity. Our setting is narrower: summarization-based compaction in long-horizon agents, where an LLM rewrites a heterogeneous source trajectory into the handoff from which the backend continues planning, tool use, or memory update. This makes compression a security boundary rather than only an efficiency layer. Summarization faithfulness work studies unsupported generated content
\cite{belem2025single,bao2025faithbench}, and CompressionAttack studies adversarial perturbations that induce semantic drift during compression
\cite{liu2025compressionattack}. \emph{Relinking} differs from both: the handoff may contain only source-attested action/value fragments and arise from ordinary summarization, yet express a new action--value binding that is not supported in the source.

\mypara{Agent Vulnerabilities and Attacks}
Prompt injection and prompt hacking show that LLM applications can be induced to follow attacker-provided instructions
\cite{perez2022ignore,schulhoff2023ignore,greshake2023not}. Agent settings amplify this risk because model outputs can trigger tool calls, state updates, data access, or workflow decisions
\cite{zhang2025agent,zhan2024injecagent,debenedetti2024agentdojo}. Closest to our construction are payload splitting~\cite{kang2024exploiting,li2024drattack}, which distribute adversarial content across the source to evade direct detection or enable later reconstruction. \emph{Relinking} differs on the source-side object: the source does not state the target task as a complete instruction, nor does it specify how to reconstruct it. It only contains separated action/value fragments, while the compressor creates the missing binding in the handoff.

\mypara{Agent Defenses}
LLM-agent defenses include input filters, guard models, safety prompts, instruction hierarchy, structured separation, and runtime policy enforcement
\cite{inan2023llama,zeng2024shieldgemma,deberta-v3-base-prompt-injection-v2,chen2025struq,chen2025meta,debenedetti2025defeating}.
These defenses usually certify one representation at a time: the source input, generated output, backend action, or runtime permission. Localization and traceback methods such as PromptLocate, AttnTrace, and Attention Tracker identify suspicious or influential source spans
\cite{jia2025promptlocate,wang2025attntrace,hung2025attention}. These methods are complementary, but \emph{relinking} is a transition-level failure: before compression, no malicious binding exists in the source; after compression, the handoff can appear grounded because its action/value fragments are source-attested. Our defense therefore audits whether each backend-facing binding in the handoff is licensed by a corresponding binding in the source, rather than scanning either representation in isolation.

\section{Conclusion}
We identified \emph{relinking}, a vulnerability where prompt compressors merge harmless fragments into malicious instructions. Our tool \Attack{} exploits this flaw to trigger unauthorized actions across LLM agents, evading existing defenses that ignore the compression boundary. Our boundary audit, KBRA, neutralizes the threat, showing that future agent designs must treat prompt compression as a critical security boundary.

\bibliographystyle{plain}

\appendices

\section{Benchmark Construction}
\label{app:benchmark-construction}

\mypara{Goal and filtering}
We adapt existing agent benchmarks to the compressed-agent threat model. A usable
sample needs four components: a clean long carrier context, a benchmark-derived
backend-relevant target action, an exact target argument, and enough textual
distance to place the action-side and value-side fragments separately. For each
benchmark we keep only rows whose objective maps to a concrete backend action and
that expose an exact target argument (account id, email, URL, file path, order
id, line id, or reservation id). We drop rows whose clean carrier already
contains the complete target instruction or leaks the exact value in a locally
actionable form, and rows whose carrier is too short for distributed fragment
placement under the same constraints used by \Attack{}.

\mypara{Final set}
The final evaluation set contains \textbf{619} benchmark-derived samples: 30 from
AgentDojo, 188 from ASB, 200 from InjecAgent, and 201 from \(\tau^2\)-Bench. Each
row stores its clean carrier, source benchmark, scenario metadata,
benchmark-derived target action, exact target argument, and entity type, so the
same \Attack{} construction and evaluation pipeline applies across all
benchmarks. AgentDojo/ASB rows are reconstructed into long clean-agent packets
with concrete backend targets; InjecAgent rows pair backend-actionable
objectives with benign carriers; \(\tau^2\)-Bench rows are kept as
task-preserving customer-service carriers without LLM summarization.

\section{RQ1 Falsification Results}
\label{app:rq1}
\subsection{Attention-Routing Statistics}
\label{app:rq1-stats}
\(\Delta_{\mathrm{attn}}\) on the Gemma-4-31B-it attention-feasible subset
(\(163\) samples). All per-benchmark \(95\%\) bootstrap CIs lie strictly above
zero.
\begin{table}[h]
\centering
\small
\setlength{\tabcolsep}{4pt}\renewcommand{\arraystretch}{1.08}
\caption{$\Delta_{\mathrm{attn}}$ statistics. Overall: sign-test
$p=2.8\!\times\!10^{-47}$, Wilcoxon $p=1.7\!\times\!10^{-28}$.}
\label{tab:rq1-attn-stats}
\begin{tabular}{@{}lrrrl@{}}
\toprule
\textbf{Source} & \textbf{Rows} & \textbf{Pos.} & \textbf{Mean} &
\textbf{$95\%$ CI} \\
\midrule
AgentDojo  & $13$ & $100.0\%$ & $0.000411$ & $[0.000256,\,0.000676]$ \\
ASB        & $50$ & $100.0\%$ & $0.001620$ & $[0.001516,\,0.001718]$ \\
InjecAgent & $50$ & $98.0\%$  & $0.001311$ & $[0.001121,\,0.001505]$ \\
$\tau^2$-bench  & $50$ & $100.0\%$ & $0.001117$ & $[0.001048,\,0.001186]$ \\
\midrule
Overall    & $163$ & $99.4\%$ & $0.001275$ & --- \\
\bottomrule
\end{tabular}
\end{table}

\subsection{Routing Direction and Long-Context Validation}
\label{app:rq1-direction}
Due to causal masking the binding signal flows exclusively in the
\(R_v\!\to\!R_p\) direction (reverse direction zero on all valid rows). Gemma-4-31B-it
full-attention extraction OOMs on the longest AgentDojo/$\tau^2$-bench contexts
(\(231\) rows at \(8\) H800 GPUs), which are measurement failures, not negative
samples; a Qwen3.5-4B probe under the identical construction confirms one-way
routing on \(96.7\%\)/\(100\%\) of these long contexts. 

\subsection{Type-Mismatch Diagnostic (Framing)}
\label{app:rq1-typemismatch}
Stricter test of Insight~2: compatible vs.\ type-mismatched construction at equal
exposure (full \(619\)-sample set). A positive paired difference means type
compatibility, not exposure, drives the framing boost.
\begin{table}[h]
\centering
\small
\setlength{\tabcolsep}{4pt}\renewcommand{\arraystretch}{1.08}
\caption{Type-mismatch paired difference (match $-$ mismatch).}
\label{tab:rq1-typemismatch}
\begin{tabular}{@{}lrrrr@{}}
\toprule
\textbf{Source} & \textbf{Rows} & \textbf{Match} & \textbf{Mismatch} &
\textbf{Diff.} \\
\midrule
AgentDojo  & $30$  & $-0.005$ & $-0.343$ & $+0.338$ \\
ASB        & $188$ & $-0.535$ & $-0.977$ & $+0.441$ \\
InjecAgent & $200$ & $\phantom{-}0.164$ & $\phantom{-}0.127$ & $+0.037$ \\
$\tau^2$-bench  & $201$ & $-0.113$ & $-0.007$ & $-0.105$ \\
\midrule
Overall    & $619$ & $-0.146$ & $-0.275$ & $+0.128$ \\
\bottomrule
\end{tabular}
\end{table}

\subsection{Head-Ablation Sweep (Routing Necessity)}
\label{app:rq1-ablation}
Top-\(k\) vs.\ count-matched random ablation (balanced \(120\)-sample subset,
Qwen3.5-4B). Top-\(64\) reduces the canonicalization-utility gap
(\(0.397\!\to\!0.079\), routing not inert), but count-matched Random-\(64\)
reduces it more (to \(-0.131\)), so the dependence is broad, not localized to top
heads. 

\subsection{Base-vs-Instruct Canonicalization}
\label{app:rq1-basevsit}
Strict discriminative test for Insight~3. RCG is the relation-compatible gain
(raw $-$ control); RSR its positive rate; Strict credits only samples that commit
to the compatible binding \emph{and} reject the type-mismatched control. The
instruct$-$base Strict gap is \(65.9\) pp; the base model's positive control mean
shows it commits to incompatible bindings indiscriminately.
\begin{table}[h]
\centering\scriptsize
\setlength{\tabcolsep}{4pt}\renewcommand{\arraystretch}{1.08}
\caption{Discriminative canonicalization, Gemma-4-31B (base) vs.\ Gemma-4-31B-it.}
\label{tab:rq1-basevsit}
\small
\begin{tabular}{@{}lrrrrrrr@{}}
\toprule
\textbf{Model} & \textbf{Raw} & \textbf{Raw$^+$} & \textbf{Ctrl} &
\textbf{Ctrl$^+$} & \textbf{RCG} & \textbf{RSR} & \textbf{Strict} \\
\midrule
Gemma-4-31B(base)     & $0.326$ & $98.9\%$ & $\phantom{-}0.139$ & $90.5\%$ & $0.187$ & $88.4\%$ & $15.7\%$ \\
Gemma-4-31B-it  & $0.440$ & $91.6\%$ & $-0.969$ & $10.8\%$ & $1.408$ & $94.5\%$ & $81.6\%$ \\
\bottomrule
\end{tabular}
\end{table}

\section{KBRA Details}
\label{app:kbra-details}

This appendix gives the canonical projection, typed normalizers, and formal
bounds for the \kbra{} defense (\Cref{subsec:kbra}).

\subsection{Canonical Projection and Typed Normalizers}
\label{app:kbra-canonical}

\mypara{Typed canonical projection \texorpdfstring{$(\Phi)$}{}}
The projection \(\Phi:\Sigma^\ast\rightarrow \mathcal{K}\cup\{\bot\}\) parses an
extracted binding \(\sigma\) into predicate, role, value, value type \(t\), and
provenance, and returns a canonical key
\(\Phi(\sigma)=\mathrm{enc}(\kappa_p, r, t, \kappa_v, \kappa_\pi)\)
only when every component is defined by a frozen normalizer; otherwise
\(\Phi(\sigma)=\bot\). The projection is deterministic, typed, idempotent, and
threshold-free, and \(\bot\) matches nothing (including itself), so uncertainty
becomes denial rather than authorization. The source and post key sets are
\(\mathcal{K}_H=\{\Phi(\sigma):\sigma\in E(H_{\mathrm{lic}}),\Phi(\sigma)\neq\bot\}\)
and the analogous \(\mathcal{K}_a\) on the post side. The typed value normalizers
are summarized in \Cref{tab:kbra-normalizers}; predicate and role normalizers
follow the same rule (schema names are canonical, source-side phrases must map
through frozen tables, unknown/incompatible phrases return \(\bot\)).

 \begin{table}[t]
  \centering
  \scriptsize
  \setlength{\tabcolsep}{3pt}
  \renewcommand{\arraystretch}{1.1}
  \caption{\kbra{} typed normalizers. Each normalizer emits a canonical key or
  \(\bot\).}
  \label{tab:kbra-normalizers}
  \begin{tabular}{@{}p{0.10\columnwidth}p{0.41\columnwidth}p{0.41\columnwidth}@{}}
  \toprule
  \textbf{Type} & \textbf{Canonicalization} & \textbf{Rejects} \\
  \midrule
  \texttt{num} &
  Parse as an exact decimal or rational; strip separators and leading zeros;
  normalize sign. &
  Malformed numbers, approximate quantities, or values requiring rounding. \\
  \midrule
  \texttt{qty} &
  Parse magnitude and unit; convert through a frozen unit table to a canonical
  dimensioned representation. &
  Unknown units, ambiguous units, or dimension mismatch with the declared role. \\
  \midrule
  \texttt{date} &
  Parse absolute instants or intervals at schema-declared granularity; encode in
  ISO-8601. &
  Relative, unanchored, or ambiguous dates. \\
  \midrule
  \texttt{entity} &
  Map a surface form to a canonical entity id through a frozen alias table. &
  Unknown aliases or multiple possible entity ids. \\
  \midrule
  \texttt{strid} &
  Apply Unicode NFKC; apply schema-declared case folding and whitespace folding;
  compare opaque identifiers byte-exact after normalization. &
  Malformed identifiers or values incompatible with the schema role. \\
  \bottomrule
  \end{tabular}
  \end{table}

\subsection{Formal Security and Utility Bounds}
\label{app:kbra-bounds}

Let \(\equiv\) denote true binding equivalence.

\begin{definition}[Collision Error]
The probability that an unsupported action binding \(\sigma_a\) accidentally
collides with a licenseable source key:
\(
\epsilon_{\mathrm{col}} = \sup_{\mathcal{A}:\mathrm{Lic}=0} \Pr [ \exists \sigma_H \in E(H_{\mathrm{lic}}) :
\Phi(\sigma_a) = \Phi(\sigma_H) \neq \bot,\ \sigma_a \not\equiv \sigma_H ].
\)
\end{definition}

\begin{definition}[Separation Error]
The probability that a benign action and its true source witness fail to meet at
the same canonical key:
\(
\epsilon_{\mathrm{sep}} = \Pr_{\sigma\equiv\sigma_H} [ \Phi(\sigma) \neq \Phi(\sigma_H) \vee \bot \in \{\Phi(\sigma), \Phi(\sigma_H)\} ].
\)
\end{definition}

\begin{definition}[Extraction Miss Error]
\(\epsilon_{\mathrm{post}}\) is the probability that \(E\) fails to recover a true
post-compression binding from the free-text compressed handoff.
\end{definition}

\mypara{Boundary authorization bound}
Assume backend actions are structured, so every executed argument binding is read
from the runtime schema (no post-side extraction miss). For an unsupported
relinked action the dispatcher accepts only if \(\Phi(\sigma_a)\in\mathcal{K}_H\);
since the action is unsupported, any acceptance implies a collision, so the
residual backend action rate is bounded by safety,
\(
BAR_{\kbra} \le \epsilon_{\mathrm{col}}.
\)
For a benign action requiring at most \(m\) bindings, a false reject occurs if any
binding pair is separated by \(\Phi\) or maps to \(\bot\); by the union bound,
\(
\Pr[\mathrm{reject}\mid \mathrm{benign}] \le m\,\epsilon_{\mathrm{sep}}.
\)

\mypara{Compressed-source asymmetry}
Auditing the compressed handoff directly relies on free-text post-side
extraction. Because \kbra{} fails closed, an extraction miss
(\(\epsilon_{\mathrm{post}}\)) yields a rejection, never an authorization, so the
safety bound is unchanged (\(BAR_{\kbra} \le \epsilon_{\mathrm{col}}\)), while the
utility bound degrades to
\(
\Pr[\mathrm{reject}\mid \mathrm{benign}] \le m\,\epsilon_{\mathrm{sep}} + \epsilon_{\mathrm{post}}.
\)
This is the tradeoff: the compressed source audits earlier but pays
\(\epsilon_{\mathrm{post}}\) in benign utility relative to the structured backend
proposal. (Only a free-text, fail-open deployment would move
\(\epsilon_{\mathrm{post}}\) into safety; our evaluation never does this.) The
exact and semantic matchers and the four evaluated modes share this bound and are
detailed in the released artifact.

\subsection{Adaptive Adversaries against KBRA}
\label{app:kbra-adaptive}

We grant the adversary full white-box knowledge of KBRA---the invariant,
the projection $\Phi$, the typed normalizers and alias table, and the
channel-to-authority mapping---and let it shape any low-authority segment
of $H_{\mathrm{untrust}}$. Two strategies follow from KBRA's two
acknowledged soft spots.

\noindent\textbf{A1: Single-segment authorized binding.}
The adversary collapses $p^{\star}$, $r^{\star}$, $v^{\star}$ into one
low-authority segment, making the binding locally supported
($\mathrm{Support}=1$) so it passes the locality requirement. This
abandons \emph{relinking}: $\tau^{\star}$ now appears verbatim in
$H_{\mathrm{adv}}$, the attack is no longer attributable to the
compression boundary, and it reduces to the direct-injection case KBRA
defers to standard input filters. A1 trades success for
\textbf{detectability}---the complete payload exists in $H$ before
compression, where input-level defenses operate. KBRA does not claim to
block A1; it makes the compression boundary no longer exploitable,
forcing the adversary back to a representation prior defenses cover.

\noindent\textbf{A2: Entity-collision.}
The adversary keeps the binding split but crafts a value engineered to
collide, under the \emph{semantic} matcher, with a legitimate source
key---exploiting the residual $\epsilon_{\mathrm{col}}$ bounded in
Appendix~\ref{app:kbra-bounds}. The locality argument is unaffected:
under the \emph{exact} matcher no collision is admitted and residual BAR
is $0.0\%$ (Table~\ref{tab:defense-security-utility}); under the \emph{semantic} matcher the
residual is upper-bounded by $\epsilon_{\mathrm{col}}$ and matches the
$0.2\%$--$0.3\%$ in Table~\ref{tab:defense-security-utility}. This residual shrinks
monotonically as the frozen alias table is expanded, isolating the
leakage to a single auditable artifact rather than to the locality
mechanism that defeats \emph{relinking}.
\end{document}